\documentclass[
    11pt,
    letterpaper,
    reprint,
    notitlepage,
    superscriptaddress,
    aps,pra
]{revtex4-2}

\UseRawInputEncoding

\usepackage{amsmath,amssymb,amsfonts} 
\usepackage[mathcal]{euscript}
\usepackage{mathtools}
\usepackage{siunitx}
\usepackage{physics}
\usepackage{microtype}
\usepackage{soul}

\usepackage{subdepth}   

\usepackage{bm} 
\renewcommand{\mathbf}{\bm}
\usepackage{dsfont} 
\renewcommand{\mathbb}{\mathds} 

\usepackage{graphicx}
\usepackage[svgnames,dvipsnames]{xcolor}
\definecolor{NewBlue}{rgb}{0.1, 0.1, 0.7}
\usepackage[colorlinks,
    linkcolor=FireBrick,
    citecolor=MediumBlue,
    urlcolor=DarkGreen]{hyperref}

\usepackage[capitalise]{cleveref}

\newcommand{\rmd}{\mathrm{d}}
\newcommand{\rme}{\mathrm{e}}
\newcommand{\rmi}{\mathrm{i}}

\renewcommand{\t}[1]{\mathrm{{#1}}}

\newcommand{\avg}{\expval}

\newcommand{\LigoMIT}{LIGO Laboratory, Massachusetts Institute of Technology,
    Cambridge, MA 02139}
\newcommand{\MechMIT}{Department of Mechanical Engineering,
    Massachusetts Institute of Technology, Cambridge, MA 02139}

\begin{document}

\title{Wave-particle duality in the measurement of gravitational radiation}

\author{Hudson A. Loughlin}
\email{hudsonl@mit.edu}
\affiliation{\LigoMIT}
\author{Germain Tobar}
\affiliation{Department of Physics, Stockholm University, SE-106 91 Stockholm, Sweden}
\author{Evan D. Hall}
\affiliation{\LigoMIT}
\author{Vivishek Sudhir}
\email{vivishek@mit.edu}
\affiliation{\LigoMIT}
\affiliation{\MechMIT}

\begin{abstract}
In a consistent description of the quantum measurement process, whether the wave or particle-like aspect of a system is revealed depends on the details of the measurement chain, and cannot be interpreted as an objective fact about the system independent of the measurement. We show precisely how this comes to be in the measurement of gravitational radiation. Whether a wave or particle-like aspect is revealed is a property of the detector employed at the end of the quantum measurement chain, rather than of the meter, such as a gravitational-wave (GW) antenna or resonant bar, used to couple the radiation to the detector. A linear detector yields no signal for radiation in a Fock state and a signal proportional to the amplitude in a coherent state --- supporting a wave-like interpretation. By contrast, the signal from a detector coupled to the meter's energy is non-zero only when the incident radiation contains at least a single graviton. Thus, conceptually simple modifications of contemporary GW antennae can reveal wave-particle duality in the measurement of gravitational radiation.
\end{abstract}

\maketitle

\section{Introduction}

A basic lesson of quantum physics is that a measurement does not reveal a pre-existing value of an
observable, rather the measurement itself manifests the realized
outcome \cite{Bohr35,Bohr48,Bell81,Red89}.
That is, the result of a measurement, and therefore its interpretation, depends on the details of the
measurement context \cite{Red89,BudCab22}.

In standard quantum mechanics, this aspect is evident in a simple model of indirect measurement of a system
using a ``meter'' and a ``detector''.
(For instance, in the Stern-Gerlach experiment, the atomic spin is the system, the atom's position is the meter, and the screen is the detector.)
Suppose a system, in state $\ket{S}$, is coupled to a meter, in state $\ket{M}$, such that their joint
initial state is $\ket{\t{SM}} = \ket{S}\ket{M}$.
Expressing the system state in a complete orthogonal basis $\{\ket{X}\}$, $\ket{S} = \sum_X S_X \ket{X}$,
let the system and meter evolve unitarily such that the resulting joint state,
$\ket{\t{SM}'} = \sum_X S_X' \ket{X}\ket{M_X}$, entails correlations between the meter and system. However, this does not constitute a measurement since the basis $\{\ket{X}\}$ is in no way preferred over any other.
The meter subsequently interacts unitarily with a detector prepared in state $\ket{D}$.
Let this interaction affect the transition $\ket{M_X} \ket{D} \rightarrow \sum_Y C_{XY} \ket{M_Y} \ket{D_Y}$.
The resulting joint state of the system, meter, and detector is $\ket{\t{SMD}'} = \sum_{XY} S^\prime_X C_{XY} \ket{X} \ket{M_Y}\ket{D_Y}$. 

The essential role of the detector is to map the quantum state of the meter, vis-a-vis
the correlations established between itself and the meter, into a distinguishable and objective output.
We suppose that $\{\ket{D_Y}\}$ is orthonormal in the detector's Hilbert space,
and therefore distinguishable~\cite{Berg10} (even if $\{\ket{M_Y}\}$ need not be). If we also assume that the detector's
quantum state is unobservable, the system and meter are left in the (mixed) state $\Tr_D \ketbra{\t{SMD}'}
= \sum_Y \abs{C_Y}^2 \ket{Y}\braket{Y}{M_Y}\bra{M_Y}$, where $\ket{Y}$ is defined by $C_Y\ket{Y}=\sum_X S_X' C_{XY}\ket{X}$.
Effectively, the system is measured in the basis $\ket{Y}$, leaving the meter in the state $\ket{M_Y}$ with probability $\abs{C_Y}^2$.
In particular, the coupling between the meter and detector determines the coefficients $C_{XY}$ and thus the measurement basis of the system.

Thus, by choice of which basis the meter is readout, measurements of entirely different observables of the system
can be realized.
This conclusion relies on (a) the validity of the quantum superposition principle for the system;
(b) entanglement between the system and meter; and, (c) the presence of a further ``detector'' which
measures the meter in a preferred basis (for example by some form of super-selection~\cite{Sud76,Peres80,Zurek81,Zurek82,Walls85})
so as to realize the objective (``classical'') outcome of the measurement.

These ideas are manifest~\cite{GranAsp86,Mandel99,Zeil99,DurRemp98,BertHar01,Shad14}
(but subtle~\cite{WiseHar95}) in the measurement of electromagnetic radiation:
individual ``clicks'' of an absorptive detector correspond to ``photons'', whereas correlations between the ``clicks''
of such detectors reveal an interference pattern ascribing a wave-like reality to the radiation.
This complementarity --- ``wave-particle duality'' --- is one signature
of the quantum character of the electromagnetic radiation.

The ability to directly detect gravitational radiation \cite{ligo16}
using GW antennae~\cite{Adhi14,ReitSaul19}, or proposals to see a quantum jump
due to its absorption by an elastic bar~\cite{TobPik24,KahTrick24,TobTob25},
again brings up the question of complementarity, now in the context of gravitational radiation. In particular,
how does the choice of detector reveal the wave or particulate character of gravitational radiation?

In this paper, we examine the response of GW transducers, including antennae and bars, undergoing continuous measurement of either a single quadrature of the meter (for example, by homodyning), or of its energy.
We find that a homodyne measurement yields a measurement record whose intensity is linearly proportional to the
amplitude of the incident gravitational radiation. In contrast, an energy measurement emits a record which contains
quantum jumps corresponding to the energy of a graviton. That is (to paraphrase
Glauber~\cite{MuthZub03}), ``a graviton is what an energy-coupled gravitational-wave-transducer detects''.

Our results extend the principle of complementarity to quantized gravitational radiation. Further, we
highlight how modifications to existing gravitational wave antennas such as replacing homodyne
measurements of the output photon flux with single-photon detectors, can be used to probe the
wave--particle duality of gravitational radiation. All this is achieved by providing a unified framework
to quantize gravitational radiation and its interaction across interferometric and bar-type transducers, and 
their subsequent readout via linear or energy-absorbing detectors.
This resolves a long-standing conceptual gap, clarifying how gravitational wave antennae or bars can function 
analogous to either the linear or energy detectors widespread in quantum optics.

\begin{figure}[t!]
	\centering
	\includegraphics[width=1\columnwidth]{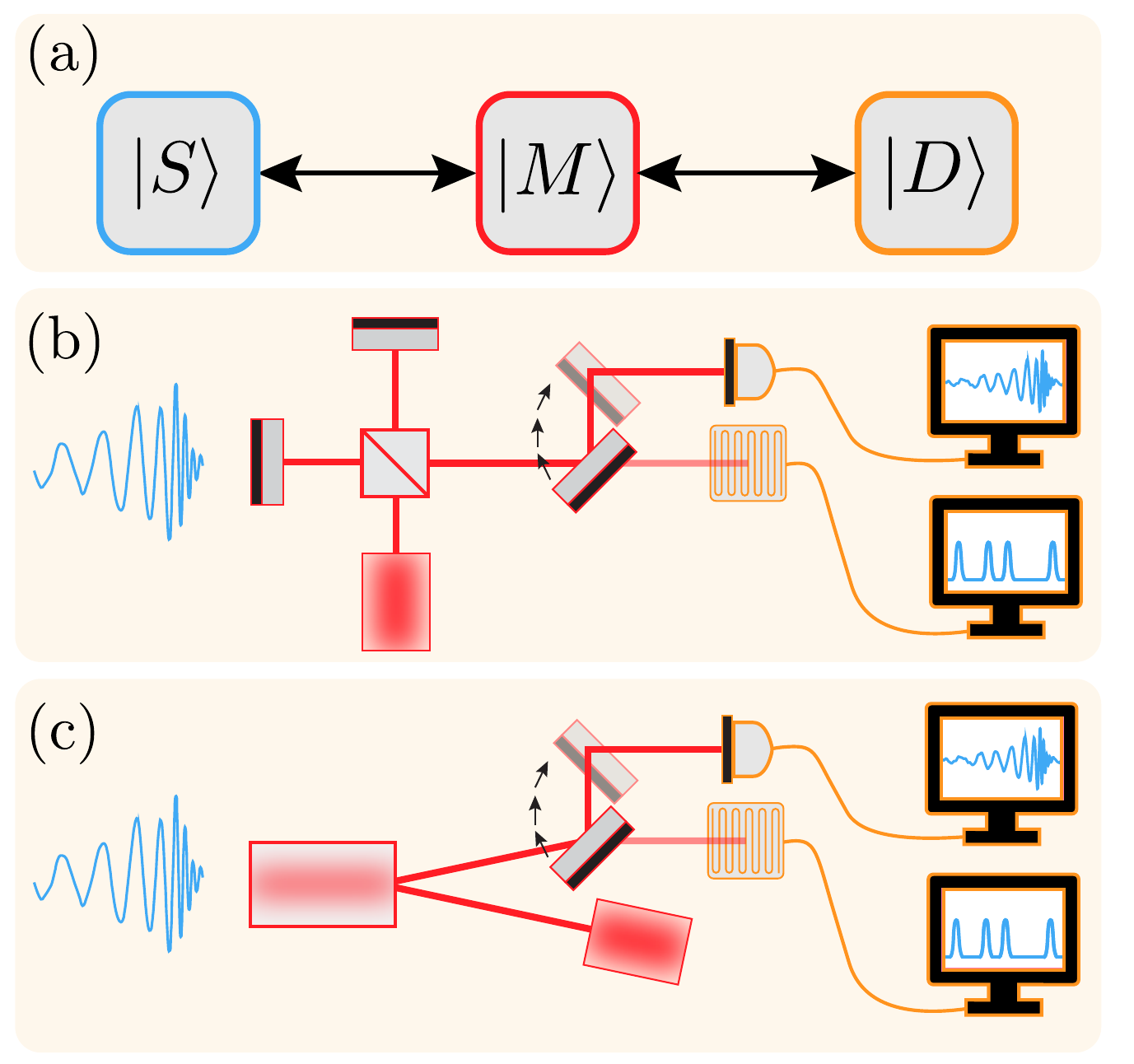}
	\caption{\label{mainfig}
	Comparison of detector responses to gravitational radiation for different measurement schemes. (a) Schematic representation of continuous quantum measurement setups with a system in state $|S\rangle$ coupled to a meter in state $|M\rangle$ and measured with a detector. The meter can be either a gravitational wave antenna, as in (b), or a bar, as in (c). Complementarity dictates that the choice of basis for the measurement of the meter determines whether the detector records a continuous signal or a stream of discrete clicks.
	}
\end{figure}

\section{Quantized gravitational wave states}

We adopt the view that gravitational radiation can be described quantum mechanically.
That is, the linearized metric $h_{\mu \nu}$ is a quantum field on a fixed Lorentzian background spacetime.
Its physical (i.e., gauge-free) degrees of freedom can be isolated in the transverse-traceless (TT) gauge;
assuming that it is fully polarized, we are left with a single degree of freedom which can be expressed 
as a superposition of quantized field modes (see \cref{quantderivation})
\begin{equation} \label{qtisedgravfield}
    \hat{h}(t)=\int\limits_{-\infty}^{\infty}
    h_{\Omega_0}\left(\hat{a}[\Omega] \mathrm{e}^{-\mathrm{i}\left(\Omega+\Omega_0\right) t}+\text {H.c.}\right) \dd \Omega,
\end{equation}
where $\Omega_0$ is the GW carrier frequency, $h_{\Omega_0} = \sqrt{16 \hbar  G/(c^3 \pi \Omega_0 A)}$  ensures normalization of the GW flux through a cross-sectional area $A$, and $\hat{a}[\Omega]$  satisfies the commutation relation
$[\hat{a}[\Omega], \hat{a}[\Omega^{\prime}]^\dagger] = 2\pi \delta[\Omega - \Omega^{\prime}]$.
This quantization scheme allows us to consider quantum states of the propagating GW in analogy with propagating electromagnetic
waves \cite{Blow90}. For example a propagating coherent state of the GW is
\begin{equation}\label{baraeq}
    \ket{\bar{a}} = \exp \left[\int \left(\bar{a}[\Omega] \hat{a}^{\dagger}[\Omega]
    -\bar{a}^*[\Omega] \hat{a}[\Omega]\right) \dd \Omega\right] \ket{0},
\end{equation}
where $\bar{a}[\Omega]$ is the Fourier transform of its mean amplitude $\bar{a}(t) = \ev*{\hat{a}(t)}{\bar{a}}$,
and $\ket{0}$ is the vacuum state of the GW.
In order to explore the response of detectors to states of the incoming GW with a definite number of gravitons, we will also consider the GW Fock state
\begin{equation}\label{nxi}
    \ket{n,\xi} = \frac{1}{\sqrt{n!}}\left[\int \xi^*[\Omega] \hat{a}^\dag[\Omega] \dd \Omega\right]^n \ket{0},
\end{equation}
where $\xi[\Omega]$ is the frequency envelope of the propagating state normalized as $\int \abs{\xi[\Omega]}^2 \dd \Omega =1$. 
For the propagating coherent states given in \cref{baraeq}, the mean strain amplitude is
$\avg*{\hat{h}[\Omega]} \approx 4\pi h_{\Omega_0} \bar{a}[\Omega]$,
assuming for simplicity that $\bar{a}$ is real and approximately symmetric about the carrier frequency. 
For the Fock states of \cref{nxi}, $\langle \hat{h}[\Omega] \rangle = 0$. For the number flux operator of the propagating gravitational field from an idealized source a distance $R$ away from the detector, the expectation value is given by $\langle \hat{N}_{\mathrm{gw}} \rangle \approx \frac{c^3 R^2 \Omega_0}{6 \hbar G} \int_{-\infty}^{+\infty} \frac{\mathrm{d} \Omega}{2 \pi} \bar{S}_{h h}[\Omega]$, where  $\bar{S}_{hh}[\Omega]$  is the power spectral density of the 
gravitational wave field (see \cref{quantderivation}). In the case of the Fock states defined in \cref{nxi}, this evaluates to be $\langle \hat{N}_{\text{gw}} \rangle = n$, which corresponds to the number-flux of propagating gravitons in the Fock state.
For a coherent state, the contribution of the quantum fluctuations in the GW field are suppressed by $\abs{\bar{a}}^{-1}$
in $\avg*{\hat{N}_\t{gw}}$; i.e. for typical sources, the mean graviton flux is insensitive to quantum fluctuations
of the GW field \cite{CarnRod24}.

\section{Gravitational radiation as the subject of measurement}

In what follows, the GW field is the subject of measurement, i.e.
the system in the terminology of quantum measurement. Irrespective of what meter it couples to, and what detector is employed to readout the meter state, the detector must emit a continuous classical record; this is precisely 
what distinguishes a detector, and demarcates the point in the measurement chain beyond which everything can be treated
classically. In a continuous measurement, the output of a detector is a random process, which can be modeled as an operator
$\hat{Y}(t)$. The condition that it is effectively classical is that \cite{BragKhal,Bela94} 
$[\hat{Y}(t),\hat{Y}(t')] = 0$ for all $t,t'$. 
This ensures that the multi-time joint probability distribution of the output can be defined unambiguously, 
and exists as a legitimate classical probability distribution. We now analyse two specific measurement
chains whose outputs are qualitatively different depending on choice of meter-detector coupling.

When a GW passes by an interferometric GW transducer (such as Advanced LIGO) the optical field stored in the interferometer
experiences fluctuations in proportion to the GW amplitude.
This can be described by the interaction Hamiltonian (in the TT gauge) \cite{PangChen18},
\begin{equation}\label{eq:HintIFO}
    \hat{H}_\t{int} = -\frac{\hbar \omega_0}{2}\bar{\alpha} \hat{\alpha}_1 \hat{h}.
\end{equation}
Here, $\omega_0$ is the optical resonance frequency, $\bar{\alpha} = 2\sqrt{(L/c)P_{\text{cav}}/\hbar\omega_0}$ is the coherent state amplitude of the optical cavity with intracavity power $P_{\text{cav}}$ and length $L$, and $\hat{\alpha}_1 =(\hat{d}+\hat{d}^{\dagger})/\sqrt{2}$ is the operator corresponding
to the linearized amplitude of the optical field. Later, we will also employ the phase quadrature $\hat{\alpha}_2 =
-i(\hat{d}-\hat{d}^{\dagger})/\sqrt{2}$, such that $\left[\hat{\alpha}_1, \hat{\alpha}_2\right]=i$.
On the other hand, the interaction between a GW and the fundamental phonon mode of a resonant bar transducer is described by
\begin{equation}\label{eq:HintBAR}
    \hat{H}_{\mathrm{int}} = \frac{M L}{\pi^2} \hat{x}  \ddot{\hat{h}}.
\end{equation}
where $M$ is the total mass of the bar of length $L$, 
$\hat{x} = x_\t{zpm} (\hat{b}+\hat{b}^\dagger)$ the displacement of its fundamental mode with
zero-point amplitude $x_\t{zpm}=\sqrt{\hbar/(M\omega_m)}$ when it oscillates at its resonance 
frequency $\omega_m$.
Typically, the interaction between GW and a resonant bar is modeled in the local 
inertial frame~\cite{Weber60}; in Sec. IV of the SI, we reconcile this with our TT gauge description of the GW. 
In sum, the interaction between a quantized GW with either an interferometric or a resonant bar transducer can be described on the same footing using the quantized GW in the TT gauge.

\section{Linear detection of gravitational radiation\,---\,sensitivity to the wave-like features}

For an interferometric transducer, the GW couples to the amplitude, therefore its conjugate, 
the phase of the optical field, is driven by the passing GW. 
The phase of the optical field leaking out of the interferometer carries
information about the GW, and can be thought of as the meter in this setting.
For a GW well inside the interferometer's detection bandwidth, $\kappa$, such that $\Omega_0 \ll \kappa$, the output phase quadrature fluctuations in the frequency domain are (see \cref{app:cavityGwAntenna})
\begin{equation}
    \hat{\alpha}_2^{\text{out}}[\Omega]=-\frac{2\hbar \bar{\alpha}^2 \omega_0^2}{L^2 m \kappa \Omega^2} \hat{\alpha}_1^{\text{in}}[\Omega] - \hat{\alpha}_2^{\text{in}}[\Omega] + \frac{\bar{\alpha} \omega_0}{\sqrt{2 \kappa}} \hat{h}[\Omega].
\end{equation}
Here, $\hat{\alpha}_{1,2}^\t{in}$ are the quantum vacuum fluctuations in the amplitude and phase of the optical field that
enter the interferometer, $\kappa$ the rate at which light leaks in and out of it,
$L$ is its length, and $m$ is the mass of the interferometer mirror (approximated as a free-mass here).
Clearly, by measuring the mean output phase of the optical field, the amplitude of the GW can be inferred, as displayed in the first row of Table~\ref{tablewavepart}. Indeed,
an optical homodyne detector effectively performs such a measurement, so that the detector's mean photocurrent output
\begin{equation}\label{hhere}
    \avg*{\hat{I}^\t{out}[\Omega]} \propto
    \langle\hat{\alpha}_2^\text{out}[\Omega]\rangle
    = \frac{\bar{\alpha} \omega_0}{\sqrt{2 \kappa}} \langle \hat{h}[\Omega] \rangle + \text{noise terms},
\end{equation}
is proportional to the amplitude of a coherent GW. In particular, if the GW is in a Fock state,
the expectation value in \cref{hhere} evaluates to zero.

Qualitatively similar behavior holds if a resonant bar is used to transduce the GW, and the 
displacement of the bar is continuously monitored \cite{Weber60,blair94,MCerdonio_1997,aguiar2011past}.
Such a measurement chain can be modeled by treating the bar as a harmonic oscillator, whose position $\hat{x}$ 
is coupled at a rate $g$ to an electromagnetic cavity field $\hat{d}$, with cavity decay rate $\kappa$, via the interaction \cite{Clerk04,AspKipMarq14} $\hbar g \hat{x} \hat{\alpha}_1$,
such that the cavity output is subjected to homodyne detection.
The analysis can be carried out as before (see \cref{app:continuousBarMeas}), with the result that the mean output photocurrent
\begin{equation} \label{homodyneoutbar}
    \langle \hat{I}^{\text{out}} [\Omega] \rangle
    = - i\frac{4 g}{\sqrt{\kappa}}\sqrt{\frac{ M}{\hbar\omega_{\text{m}}}}
    \frac{2 L\Omega_0^2}{\gamma_{\text{m}} \pi^2} \langle \hat{h}[\Omega] \rangle + \text{noise terms},
\end{equation}
is again linear in the GW strain, with mechanical decay rate $\gamma_{\text{m}}$, and qualitatively similar to the output of an interferometric GW transducer
readout via homodyning [\cref{hhere}], as also displayed in the second row of \cref{tablewavepart}.
We assume that the gravitational wave frequency is very near the bar's resonant frequency such that $|\omega_{\text{m}} - \Omega_0| \ll \gamma_{\text{m}}$.
By design, gravitational wave antennas have much broader bandwidths than resonant bar transducers, so more astrophysical gravitational waves will satisfy this criterion for antennas than for bars.

In both cases, the measurement chain responds to a GW field of arbitrarily low
amplitude, even when the GW field cannot be described as containing even a single graviton. Indeed, the output
photocurrent, being linear in the GW strain, contains an imprint of the wave-like features of the GW field.
Further, in both cases, if the GW field is in a Fock state, the expected value of the photocurrent record is identically zero.
In this sense, all current GW detectors\,---\,which includes the totality of the transducer and the
subsequent quantum measurement chain\,---\,are only sensitive to the wave characteristics of the GW radiation.

\begin{table}[t]
\begin{ruledtabular}
\begin{tabular}{lccc}
Detection & \text { Field state } & \text { Mean Response }  \\
\hline
\text {LIGO homodyne} & $\ket{\bar{a}}$ & $\sqrt{\eta_\mathrm{ifo}\left[\Omega_0\right]}\bar{a}$     
\\
\text {Bar homodyne}  & $\ket{\bar{a}}$  & $\sqrt{\eta_\mathrm{bar}\left[\Omega_0\right]}\bar{a} $    
\\
\text {LIGO absorptive} & $\ket{\bar{a}}$ & $\eta_\mathrm{ifo}\left[\Omega_0\right]\bar{a}^2$    
\\
\text {Bar absorptive} & $\ket{\bar{a}}$ &$\eta_{\mathrm{bar}}\left[\Omega_0\right] \bar{a}^2$ 
\\
\text{LIGO homodyne} & $\ket{n}$ & 0    \\
\text{Bar homodyne}  & $\ket{n}$ & 0    \\
\text{LIGO absorptive} & $\ket{n}$ & $\eta_\mathrm{ifo}\left[\Omega_0\right]n$   \\
\text{Bar absorptive} &   $\ket{n}$ & $\eta_{\mathrm{bar}}\left[\Omega_0\right] n$ 
\\
\end{tabular}
\end{ruledtabular}
\caption{\label{tablewavepart} Scaling of the response of the detector for either coherent field states of amplitude $\bar{a}$ or Fock states with occupation number $n$. A homodyne measurement of the transducer has a response to the gravitational radiation field that is linearly proportional to the amplitude of a coherent state $\lvert\bar{a}\rangle$, but insensitive to the occupation number of a Fock state $\lvert n \rangle$. An energy measurement functions as a square-law detector that clicks only when the detector is exposed to at least a single graviton. These complementary wave and particle aspects of the field can be probed in existing gravitational wave transducers, either bar transducers or interferometric (``ifo''), simply by modifying the observable that is measured via the detector. The response is scaled by the efficiency of either the bar transducer $\eta_{\mathrm{bar}}$ or the interferometric transducer $\eta_{\mathrm{ifo}}$.}
\end{table}

\section{Energy detection of gravitational radiation\,---\,counting gravitons}

We now examine an ``energy'' detector, in which the particle number of the meter is read out and not its quadrature.
For the case of interferometric GW transducers (such as LIGO), this involves directing the optical field leaking
out of the interferometer onto a single-photon detector (after filtering out the coherent
carrier~\cite{CohPain15,HongGrob17,VelGall19,GalPolz20,gquest25}) or reducing the local oscillator strength in a sufficiently low noise homodyne detector to the point that it becomes a single-photon detector \cite{zhang12}. The resulting click rate (see \cref{app:cavityGwAntenna})
\begin{equation}
     \langle\hat{N}^\text{out}(t)\rangle =
        \frac{\bar{\alpha}^2 \omega_0^2}{4 \kappa} \int\limits_{-\infty}^\infty \frac{\text{d} \Omega}{2 \pi} \bar{S}_{h h}[\Omega] + \text{noise terms},
\end{equation}
is directly sensitive to the power spectrum, $\bar{S}_{hh}$, of the GW field, and is therefore non-zero for the GW field
in a Fock state, as summarized in rows~3 and~7 of \cref{tablewavepart}. Thus, modifying \emph{only} the output of current interferometric GW transducers can make them
graviton counters. 
 A resonant bar transducer can also be converted into a graviton counter. This can be done by dispersively coupling its phonon number $\hat{b}^\dagger \hat{b}$ to an electromagnetic cavity field, i.e. an interaction of the form $i \hbar B\, \hat{b}^\dagger \hat{b} (\hat{d}^\dagger -\hat{d})$, as proposed in Ref.~\cite{TobPik24}. Alternatively, subjecting the field leaking out of this readout cavity to photon counting (which is most practically implemented by coupling the bar to a series of successively smaller oscillators, with the field leaking from the final element directed to a detector~\cite{TobTob25}) results in a click rate (see \cref{app:continuousBarMeas})
\begin{equation}\label{nout1}
\begin{split}
    \langle\hat{N}^\text{out}(t)\rangle &= \frac{16 g^2 M L^2 \Omega_0^3}{\pi^4 \hbar \kappa \gamma_m^2} \int\limits_{-\infty}^\infty \frac{\text{d} \Omega}{2 \pi} \bar{S}_{h h}[\Omega] + \text{noise terms.}
\end{split}
\end{equation}
Just as with photon counting at the output of an interferometric GW transducer, the output here is non-zero for a
Fock state of the GW field, and is in fact proportional to the number of gravitons, as summarized in rows~4 and~8 of \cref{tablewavepart}.

In fact, in both cases, a GW field in a coherent state of amplitude $\bar{a}$, leads to a mean detector output
$\avg*{\hat{N}^\t{out}} \propto \abs{\bar{a}}^2$; i.e., proportional to the square of the coherent state amplitude,
which is the mean number of gravitons.
However, the \emph{key and qualitative difference} to a linear detector is that an exceptionally weak
coherent state with $\bar{a} \ll 1$, such that the field cannot be described as consisting of even a 
single graviton, results in an exponentially suppressed click rate. 
This can be seen in the wait-time distribution $w(\tau\mid t)$ which gives the probability that, 
given a ``click'' has occurred at time $t$, another ``click'' occurs within the succeeding interval $\tau$.
A straightforward extension of techniques from photo-detection \cite{Carm93} allows us to compute the
wait-time distribution for a graviton counter (see \cref{app:waitTime})
\begin{equation}\label{wmid}
    w(\tau \mid t)=\frac{\eta}{4}|\bar{a}|^2 \exp \left(-\tau \frac{\eta}{4}|\bar{a}|^2\right).
\end{equation}
Here $\eta$ is the efficiency of the detector relative to the GW flux intercepted by it.
Clearly, for a weak coherent GW field that does not contain even a single graviton on average,
the wait-time probability goes to zero; i.e., the wait time becomes infinite. This is true even for a high efficiency
($\eta \sim 1$) detector.
In this sense a graviton detector does not click unless the field contains at least a single graviton
(i.e., $\abs{\bar{a}} \gtrsim 1$), whereas a linear detector will still produce a weak current linearly 
proportional to the GW amplitude.

Since measurements capable of resolving non-classical states of GWs are possible in principle, it is worth entertaining
the possibility of quantum state tomography of GWs \cite{CarnRod24}. However, the efficiency of a graviton counter, 
relative to the source, is abysmally small; for example, for a modified interferometric or bar
transducer, the efficiency is
\begin{equation}
\begin{split}
    \eta_{\text{ifo}}[\Omega_0] &= \frac{3 \hbar G \kappa \bar{\alpha}^2 \omega_0^2}{c^3 R^2 \Omega_0 (\kappa^2 + \Omega_0^2)} 
        \sim 10^{-73},\\
    \eta_{\text {bar }}\left[\Omega_0\right] &= \frac{96 g^2}{\kappa} \frac{M L^2 \Omega_0^3 G}{\pi^3 c^3 \omega_{\text{m}} \gamma_{\text{m}}^2 R^2}
        \sim 10^{-61},
\end{split}
\end{equation}
for typical astrophysical sources such as the ones observed by contemporary interferometric 
GW transducers \cite{gwtc3} (see \cref{app:cavityGwAntenna,app:continuousBarMeas}). 
Even with these low quantum efficiencies, coarse quantum features of GW radiation, such as complementarity, can be observed.
For example, with current GW transduction technology, and assuming a GW strain of $10^{-22}$, we have 
$\eta |\bar{a}|^2 \gtrsim 10^6$, so it is possible to observe the wave and particle aspects of GWs with known sources
and current transducers, by only modifying how the transducer is read-out.
Although in principle it is possible to observe quantum features of a quantum field using high efficiency 
linear detectors (such as the homodyne detectors routinely used to characterize optical squeezing), 
the efficencies of GW transducers make it unviable to use them in a linear measurement chain 
to observe any non-classical features of the GW
field \cite{ParWilZah21,manikandanWilczekcomp,CarnRod24}. 


The primary experimental challenge in implementing click-based graviton detection with bar transducers 
lies in extending single phonon detection techniques, demonstrated for 
nanomechanical resonators \cite{lecocq_resolving_2015,ArranSaf19,MaLeh21}, to the end mass 
of multi-mode elastic bars \cite{TobTob25}. 
For intereferometric detection, the primary experimental challenge is to extend optical filtering 
of the sideband for single photon detection \cite{CohPain15,HongGrob17,VelGall19,GalPolz20} 
to the acoustic frequency regime relevant to today's
interferometric gravitational wave transducers.

\section{Conclusion}

From the unambiguous vantage point of an objective measurement record, we can safely assert that whether the
record reveals a wave-like or particulate character of gravitational waves depends on how the GW is itself measured.
The outcome of linear measurements produces a wave-like signal whereas an energy measurement produces a particle-like
signal. That is, the measurement of gravitational radiation, assuming it is quantized, exhibits 
wave-particle duality exactly as in the measuerement of quantized electromagnetic
radiation. These results extend the quantum mechanical principle of complementarity to the measurement of 
quantized gravitational waves.
Contemporaneous GW measurements produce a wave-like signal;
our findings suggest that conceptually minor modifications to existing GW transducers could enable them to 
probe complementarity in the measurement of gravitational radiation. 

It is perhaps unsurprising that a consistent quantum mechanical description of the GW and of the measurement process implies
complementarity. 
More important is the question of what an observation of complementarity implies for the nature of gravity.
This requires refinement on two fronts: (i) an understanding of whether experimental evidence of complementarity
implies incompatibility of observables, or contextuality of measurement outcomes; 
(ii) determining, in an agnostic view of the nature of the gravitational field, 
such as using generalized probabilistic theories (GPT)~\cite{Plav23}, what GPT is consistent with the observed 
experimental evidence.
Some relevant results exist for finite-dimensional GPTs~\cite{Wolf09,BenLov22,Plav16,Schmid21}, 
but their extension to the gravitational field remains an open problem.

\section*{Acknowledgments}

H.A.L. and E.D.H. gratefully acknowledge the support of the National Science Foundation through the LIGO operations cooperative agreement PHY--18671764464. 
G.T. acknowledges support from the General Sir John Monash Foundation. 
V.S. is partly supported by an NSF CAREER award (PHY--2441238). 

\appendix

\section{Definitions and conventions}\label{app:defAndConventions}

There are several basic conventions we adopt in this work:
\begin{itemize}
    \item For the Minkowski metric, we work with the $(-,+,+,+)$ sign convention~\cite{Zee:2013dea,carroll2019}.
    \item For the Fourier transform, we adopt the convention (for $\Omega \in \mathbb{R}$)
        \begin{equation}\label{def:FT}
            \hat{X}[\Omega] = \int\limits_{-\infty}^{\infty} \rmd t \, \hat{X}(t) \, \rme^{\rmi\Omega t},
        \end{equation}
        with the inverse Fourier transform given by
        \begin{equation}
            \hat{X}(t) = \int\limits_{-\infty}^{\infty} \frac{\rmd\Omega}{2\pi} \, \hat{X}[\Omega] \rme^{-\rmi\Omega t}.
        \end{equation}
    \item The power spectral density is then taken to be defined by~\cite{ClerkRMP10}
        \begin{equation}
            S_{XX}[\Omega] = \int\limits_{-\infty}^{\infty} \rmd{t} \, \rme^{\rmi\Omega t} 
                \langle \hat{X}(t) \hat{X}(0) \rangle
        \end{equation}
        from which the symmetrized spectral density is defined as
        \begin{equation}
            \bar{S}_{XX}[\Omega] = \frac{1}{2}\left(S_{XX}[-\Omega] + S_{XX}[\Omega]\right).
        \end{equation}
\end{itemize}


\section{Quantization of Linearized Gravitational Radiation}\label{quantderivation}

Here we present our approach to quantization of the linearized gravitational field that we will use throughout the work: quantization of the first order metric perturbation of the gravitational field in the transverse traceless (TT) gauge. As we will demonstrate in later sections, this TT gauge quantization of the linearized gravitational field can be used to account for the interaction of both interferometric detectors and resonant mass detectors with quantized gravitational radiation. We assume that the gravitational radiation interacting with our detectors is in the weak-field regime, where the metric $g_{\mu \nu}$ is given by
\begin{equation}\label{gmu}
    g_{\mu \nu} = \eta_{\mu \nu} + h_{\mu \nu},
\end{equation}
where $\eta_{\mu \nu} = \operatorname{diag}(-1,+1,+1,+1)$ is the Minkowski metric, $x_\nu = (x_0,x_1,x_2,x_3) = (-ct,x,y,z)$,and $h_{\mu \nu}$ with $|h_{\mu \nu}| \ll |\eta_{\mu \nu}|$ is a weak metric perturbation.
$g_{\mu \nu}$ is a symmetric tensor, so $h_{\mu \nu}$ must be symmetric as well.

\subsection{Classical treatment}\label{classict}

Our starting point for deriving the dynamics is the Einstein--Hilbert action:
\begin{equation}\label{seh}
    \mathcal{S}_{\text{EH}}[g] = -\frac{c^4}{16 \pi G} \int\limits_M \rmd^4 x \, \sqrt{-g} \, R,
\end{equation}
with $R$ being the scalar curvature, and the prefactor ensures that extremizing this action correctly reproduces the Einstein field equations (e.g., Ref.~\cite[\S{}VI.5]{Zee:2013dea}).
Expanding the metric as in \cref{gmu}, and simplifying for the harmonic gauge $\partial_\mu h^{\mu \nu}=0$ up to $\mathcal{O}(h^2)$, 
the action reduces to
\begin{equation}
    \mathcal{S}_{\text{EH}} = -\frac{c^4}{32 \pi G} \int \rmd^4 x\left[\frac{1}{2} \partial_\mu h_{\alpha \beta} \partial^\mu h^{\alpha \beta}-\frac{1}{4} \partial_\mu h \partial^\mu h\right].
\end{equation}
The harmonic gauge does not eliminate all gauge freedom. Therefore, we further choose to work in the TT gauge such that the metric perturbation satisfies the following further gauge-fixing conditions (see \S{7.4} of Ref.~\cite{carroll2019})
\begin{equation}
\begin{split}
    h_{0 \nu} &= 0 \\
    \eta^{\mu \nu} h_{\mu \nu} &= 0.
\end{split}
\end{equation}
These further restrictions reduce the action to 
\begin{equation}\label{sehnew}
    \mathcal{S}_{\text{EH}} =-\frac{c^4}{32 \pi G} \int \rmd^4{x} \left[\frac{1}{2} \partial_\mu h_{ij} \partial^\mu h^{ij}\right],
\end{equation}
which gives the Lagrangian density 
\begin{equation}
\begin{split}
   \mathcal{L} &= -\frac{c^4}{64 \pi G}\left[\partial_\mu h_{ij} \partial^\mu h^{ij}\right] \\
   &= -\frac{c^4}{64\pi G} \left[ -\frac{1}{c^2} (\dot{h}_{ij})^2 + (\nabla_s h_{ij})^2\right],
\end{split}
\end{equation}
where we have separated $\partial_{\mu} h_{ij}$ into its temporal component $\dot{h}_{ij} = \partial_0 h_{ij}$ and its spatial component $\nabla_s h_{ij}$ for $s \in \{1,2,3\}$.
We can now compute the canonical momenta:
\begin{equation}\label{canonicij}
    \Pi_{ij} = \frac{\partial\mathcal{L}}{\partial \dot{h}_{ij}} = \frac{c^2}{32 \pi G}\dot{h}_{ij},
\end{equation}
with Poisson brackets that are computed through functional derivatives to be 
\begin{equation}\label{poisson1}
    \{h_{ij}(\mathbf{r}), \Pi_{kl}(\mathbf{r}')\} = \delta_{ik}\delta_{jl}\delta^3(\mathbf{r} - \mathbf{r}'),
\end{equation}
following from the definition of the Poisson bracket.
After a Legendre transform, the Hamiltonian density is found to be
\begin{equation}\label{H112}
  \mathcal{H} = \frac{16 \pi G \Pi^2}{c^2}+\frac{c^4(\nabla h)^2}{64 \pi G}.
\end{equation}
To convert from a canonical description of free-space radiation to propagating radiation in a single direction, we convert to the total energy of propagating gravitational radiation through a transverse cross-sectional area $A$:%
\footnote{For the particular solution of a plane gravitational wave $h_{ij}(t) = h_{ij}^{(0)} \cos[\Omega(z/c-t)]$, which satisfies $\dot{h}_{ij}(t) = -c\nabla_z h_{ij}(t)$, the Hamiltonian \cref{H112p} correctly predicts the average power flux as $(1/A)\overline{\rmd H_{\text{GW}} / \rmd t} = (c^3/32\pi G) \overline{\dot{h}_{ij} \dot{h}^{ij}}$, where the overline indicates averaging over the cycle of the wave~\cite[\S{}3.3.2]{Creighton:2011zz}.}
\begin{equation}\label{H112p}
  H_{\text{GW}}  =\frac{16 \pi G A}{c}\int \rmd{t} \; \Pi(t)^2 + \frac{c^5A}{64 \pi G} \int \rmd{t} \; (\nabla h(t))^2.
\end{equation}
Restricting the field to a finite transverse area $A$ requires
\begin{equation}
    \delta^3\left(\mathbf{r}-\mathbf{r}^{\prime}\right) \rightarrow \sum_{\mathbf{k}_\perp} \frac{\delta\left(z-z^{\prime}\right)}{A}  e^{i \mathbf{k}_\perp \cdot\left(\mathbf{r}_\perp-\mathbf{r}_\perp^{\prime}\right)}
\end{equation} 
where $\mathbf{k}_\perp$ is the wavevector component in the transverse direction and $\mathbf{r}_\perp$ the transverse position co-ordinate. Henceforth, we consider the direction of propagation of gravitational radiation to be the $z$ direction without loss of generality (therefore setting $\mathbf{k}_\perp = 0$), and therefore the $x$ and $y$ directions are transverse to the direction of propagation. This modifies the Poisson brackets to be
\begin{equation}
    \left\{h_{i j}(z), \Pi_{k l}\left(z^{\prime}\right)\right\}= \delta_{i k} \delta_{j l} \frac{\delta^3\left(z-z^{\prime}\right)}{A}.
\end{equation} 
In order to simplify the conjugate momentum and metric perturbation into their standard plane-wave forms in the TT gauge (to ultimately provide a canonical description of the energy density of a plane wave), we now consider that the action in \cref{sehnew} implies that the metric perturbation satisfies the equation of motion
\begin{equation}\label{eq:gwWaveEq1}
    \Box h_{\mu \nu} = 0.
\end{equation}
We recognize \cref{eq:gwWaveEq1} as a wave equation. For a gravitational wave propagating along the $z$ axis, it admits solutions of the form
\begin{equation}
    h_{\mu \nu} = C_{\mu \nu} e^{i (\Omega z/c - \Omega t)} + \text{c.c.},
\end{equation}
where $C_{\mu \nu}$ is a constant tensor characterizing the gravitational wave's amplitude, phase, and polarization, and $k = \Omega/c$ is the wavenumber. Henceforth, we will denote $h_+$ and $h_\times$ corresponding to the respective amplitude of each of the independent polarizations of the TT gauge. We denote polarization tensors as $e^+_{\mu \nu}$ and $e^\times_{\mu \nu}$, corresponding to each of the independent symmetric polarizations orthogonal to the direction of propagation, such that $r^i r^j e^\ell_{ij} = 0$, for $\ell \in \{+,\times \}$. For a wave propagating in the $z$ direction, the polarization tensors reduce to
\begin{equation}
    e^+_{\mu \nu} := \begin{bmatrix}
        0 & 0        & 0        & 0 \\
        0 & 1      & 0        & 0 \\
        0 & 0        & -1      & 0 \\
        0 & 0        & 0        & 0
    \end{bmatrix}
    \quad \text{ and } \quad 
    e^\times_{\mu \nu} := \begin{bmatrix}
        0 & 0        & 0        & 0 \\
        0 & 0        & 1 & 0 \\
        0 & 1 & 0        & 0 \\
        0 & 0        & 0        & 0
    \end{bmatrix}.
\end{equation}
We now have 
\begin{equation}
    h_{\mu \nu}(t) = \left( h_+ e^+_{\mu \nu} + h_\times e^\times_{\mu \nu} \right) e^{i(kz - \Omega t)} + \text{cc}.
\end{equation}
As these plane wave solutions are independent, and since $h_{\mu\nu}(t)$ is real, a general metric perturbation propagating along the axis of the vector $z$ can be written as
\begin{equation}
\begin{split}
    h_{\mu \nu}(t) = \sum_{\ell \in \{+,\times \}} \int\limits_0^\infty \frac{\rmd\Omega}{2\pi} \Big( & h_{\ell,\Omega} e^\ell_{\mu \nu} e^{i (k z - \Omega t)}
    \\ &+ h_{\ell,\Omega}^* e^\ell_{\mu \nu} e^{-i (k z - \Omega t)} \Big),
\end{split}
\end{equation}
where $h_{\ell,\Omega}$ is a complex-valued function of $\Omega$. We write this $\Omega$ as a subscript to distinguish it from a Fourier transform variable.
From now on, we will consider a gravitational-wave detector that is sensitive to only a linear combination of the two gravitational-wave polarizations; the detector's geometry defines an antenna tensor $D_{ij}$ on the spacelike indices, which defines the scalar strain $h(t) = D^{ij} h_{ij}(t)$ that the detector is sensitive to.
Thus we will consider
\begin{equation}
    \label{hmunu}
    h(t) = \int\limits_0^\infty \frac{\rmd\Omega}{2\pi} \left(  h_\Omega e^{i (k z - \Omega t)} + h_\Omega^* e^{-i (k z - \Omega t)} \right).
\end{equation}

\subsection{Quantization}\label{quant1}

We now proceed to quantize these plane-wave solutions by promoting the Poisson brackets of the classical theory to the equal time canonical commutation relations
\begin{equation}\label{hzz'}
    \left[\hat{h}(z), \hat{\Pi}(z')\right]= i\hbar \frac{\delta\left(z - z^{\prime}\right)}{A}.
\end{equation}
To do this, we promote the Fourier amplitudes in \cref{hmunu} to operators, such that $h_\Omega \rightarrow \hat{h}_\Omega$ and define the annihilation operators $\hat{a}_\Omega$ by
\begin{equation}
    \hat{h}_\Omega  = \sqrt{\frac{16 \pi \hbar G}{c^3 \Omega A}}\hat{a}_\Omega.
\end{equation}
In terms of these annihilation operators, \cref{hmunu} becomes
\begin{equation}\label{hAnnihilation}
    \hat{h}(z,t) = \int\limits_0^\infty \frac{\rmd\Omega}{2\pi} \sqrt{\frac{16 \pi \hbar G}{c^3 \Omega A}} \left(  \hat{a}_\Omega e^{i (k z - \Omega t)} + \hat{a}_\Omega^\dagger e^{-i (k z - \Omega t)} \right).
\end{equation}

Using \cref{canonicij}, the conjugate momentum is given in terms of these annihilation operators by
\begin{equation}\label{piAnnihilation}
    \hat{\Pi}(z,t) = \int\limits_0^\infty \frac{\rmd\Omega}{2\pi} i  \sqrt{\frac{c \hbar \Omega}{64 \pi G A}} \left( \hat{a}_\Omega^\dagger e^{-i(kz-\Omega t)} - \hat{a}_\Omega e^{i(kz-\Omega t)} \right).
\end{equation}

We find that the equal time commutation relations between the metric perturbation and its conjugate momentum in \cref{hzz'} are recovered if the annihilation operators obey the commutation relations
\begin{equation}
\begin{split}
    \left[ \hat{a}_\Omega, \hat{a}_{\Omega^\prime}^\dagger \right] &= 2\pi \, \delta(\Omega - \Omega^\prime) \\
    \left[ \hat{a}_\Omega, \hat{a}_{\Omega^\prime} \right] &= 0 \\
    \left[ \hat{a}_\Omega^\dagger, \hat{a}_{\Omega^\prime}^\dagger \right] &= 0.
\end{split}
\end{equation}
In this case, we can compute (setting $t=0$ for simplicity)
\begin{equation}
\begin{split}
    &\left[\hat{h}(z,t=0), \hat{\Pi}(z',t=0)\right]  \\
    &=  \frac{i \hbar}{2cA} \int\limits_0^\infty \frac{\mathrm{d} \Omega}{2\pi} \int\limits_0^\infty \frac{\mathrm{d} \Omega^\prime}{2\pi} \Big(\left[\hat{a}_\Omega, \hat{a}_{\Omega^\prime}^\dagger \right] e^{i(kz-k^\prime z^\prime)} \\
    & \qquad \qquad \qquad \qquad \qquad  - \left[\hat{a}_\Omega^\dagger, \hat{a}_{\Omega^\prime} \right] e^{-i(kz-k^\prime z^\prime)}\Big) \\
    &= \frac{i \hbar}{2A} \int\limits_{-\infty}^\infty \frac{\mathrm{d} k}{2\pi} \int\limits_{-\infty}^\infty \frac{\mathrm{d} k^\prime}{2\pi} \, 2 \pi\,  \delta(k - k^\prime) \Big(e^{i(kz-k^\prime z^\prime)} \\
    & \qquad \qquad \qquad \qquad \qquad \qquad  \qquad \quad + e^{-i(kz-k^\prime z^\prime)} \Big) \\
    &= \frac{i \hbar}{2A} \int\limits_{-\infty}^\infty \frac{\mathrm{d} k}{2\pi} \left(e^{ik(z-z^\prime)} + e^{-ik(z-z^\prime)} \right) \\
    &= i \hbar \frac{\delta(z-z^\prime)}{A} .
\end{split}
\end{equation}
Between the first and second lines, we have used the relation $\Omega= c |k|$ and the delta function identity $\delta(ax) = \delta(x)/|a|$ for $a \in \mathbb{R}$.

We can also compute the quantized Hamiltonian in terms of the annihilation operators. First, from \cref{hAnnihilation}, we compute
\begin{equation}\label{gradh}
\begin{split}
    \nabla \hat{h}(z,t) = \int\limits_0^\infty \frac{\rmd\Omega}{2\pi} i  \sqrt{\frac{16 \pi \hbar G \Omega}{c^5 A}} \Big( & \hat{a}_\Omega e^{i (k z - \Omega t)} \\
    & - \hat{a}_\Omega^\dagger e^{-i (k z - \Omega t)} \Big),
\end{split}
\end{equation}
using $k = \Omega/c$.
From \cref{H112p,piAnnihilation,gradh}, we find (taking $z=0$ for simplicity)
\begin{equation}
\begin{split}
    \hat{H}_\text{GW} &= \frac{16 \pi G A}{c} \int \rmd{t} \, \hat{\Pi}(t)^2 + \frac{c^5 A}{64 \pi G} \int \rmd{t} \, \left(\nabla \hat{h}(t)\right)^2 \\
    &= -\frac{1}{2} \int\limits_{-\infty}^\infty \mathrm{d}t \left( \int\limits_0^\infty \frac{\rmd\Omega}{2\pi} \sqrt{\hbar \Omega} \left( \hat{a}_\Omega e^{- i\Omega t} - \hat{a}_\Omega^\dagger e^{i\Omega t} \right)\right)^2 \\
    &= \frac{1}{2} \int\limits_0^\infty \frac{\rmd\Omega}{2\pi} \hbar \Omega   \left(\hat{a}_\Omega^\dagger \hat{a}_\Omega + \hat{a}_\Omega \hat{a}_\Omega^\dagger)  \right) \\
    &= \int\limits_0^\infty \frac{\rmd\Omega}{2\pi} \hbar \Omega \, \hat{a}_\Omega^\dagger \hat{a}_\Omega + (\text{zero point energy})
\end{split}
\end{equation}
We point out that the co-rotating terms $\hat{a}_\Omega \hat{a}_{\Omega^\prime}$ and $\hat{a}_\Omega^\dagger \hat{a}_{\Omega^\prime}^\dagger$ drop out since the integration region is over all positive frequencies, such that $\delta(\Omega + \Omega^\prime)$ vanishes everywhere in the integration region.

We are principally interested in gravitational waves with a carrier frequency $\Omega_0$ and quantum fluctuations on top of this mean field. We thus want to move from the lab frame we have been considering so far to the modulation frame at frequency $\Omega_0$. To do so, we define~\cite{Danilishin2012}
\begin{equation}
\begin{split}
    \hat{h}^{(+)} &= \int\limits_0^\infty \frac{\rmd\Omega}{2\pi} \sqrt{\frac{8 \pi \hbar G}{c^3 \Omega A}} \hat{a}_\Omega e^{-i \Omega t} \\
    \hat{h}^{(-)} &= \left( \hat{h}^{(+)} \right)^\dagger = \int\limits_0^\infty \frac{\rmd\Omega}{2\pi} \sqrt{\frac{8 \pi \hbar G}{c^3 \Omega A}} \hat{a}_\Omega^\dagger e^{i \Omega t}.
\end{split}
\end{equation}
We further define the operator $\hat{a}[\Omega]$ by
\begin{equation}
    \hat{a}[\Omega] := \hat{a}_{\Omega_0 + \Omega},
\end{equation}
which will turn out to be the the Fourier transform of $\hat{a}(t)$ in the modulation frame. This operator and its adjoint obey the commutation relation
\begin{equation}
    \left[\hat{a}[\Omega], \hat{a}[\Omega^\prime]^\dagger \right] = \left[\hat{a}_{\Omega_0 + \Omega}, \hat{a}_{\Omega_0 + \Omega^\prime}^\dagger \right] = 2 \pi \delta[\Omega - \Omega^\prime].
\end{equation}
With this definition, we have 
\begin{equation}
\begin{split}
    \hat{h}^{(+)} &= \sqrt{\frac{8 \pi \hbar G}{c^3 \Omega_0 A}} e^{-i \Omega_0 t} \int\limits_{-\Omega_0}^\infty \frac{\rmd\Omega}{2\pi} \sqrt{\frac{\Omega_0 + \Omega}{\Omega_0}} \hat{a}[\Omega] e^{-i \Omega t} \\
    \hat{h}^{(-)} &= \sqrt{\frac{8 \pi \hbar G}{c^3 \Omega_0 A}} e^{i \Omega_0 t} \int\limits_{-\infty}^{\Omega_0} \frac{\rmd\Omega}{2\pi} \sqrt{\frac{\Omega_0 - \Omega}{\Omega_0}} \hat{a}[-\Omega]^\dagger e^{-i \Omega t}.
\end{split}
\end{equation}
We now assume that the carrier frequency is large compared to the modulation frequency for any frequencies of interest; i.e., $\Omega_0 \gg \Omega$ for all relevant $\Omega$, such that 
\begin{equation}
\label{eq:gravitonQuantumOpticsEqs}
\begin{split}
    \hat{h}^{(+)} &\approx \sqrt{\frac{8 \pi \hbar G}{c^3 \Omega_0 A}} e^{-i \Omega_0 t} \int\limits_{-\infty}^\infty \frac{\rmd\Omega}{2\pi} \hat{a}[\Omega] e^{-i \Omega t} \\
    &= \sqrt{\frac{8 \pi \hbar G}{c^3 \Omega_0 A}} e^{-i \Omega_0 t} \hat{a}(t) 
\end{split}
\end{equation}
\begin{equation}
\begin{split}
        \hat{h}^{(-)} &\approx \sqrt{\frac{8 \pi \hbar G}{c^3 \Omega_0 A}} e^{i \Omega_0 t} \int\limits_{-\infty}^\infty \frac{\rmd\Omega}{2\pi} \hat{a}[-\Omega]^\dagger e^{-i \Omega t} \\
        &= \sqrt{\frac{8 \pi \hbar G}{c^3 \Omega_0 A}} e^{i \Omega_0 t} \hat{a}^\dagger(t).
\end{split}
\end{equation}
The assumption leading to these equations is standard in quantum optics\,---\,see, for instance, Refs.~\cite{Blow90,Danilishin2012}. In quantum optics, it is well justified since we are then generally interested in measuring optical signals with photodetectors and the maximum observable modulation frequency is limited by the electronics used in the photodetection process. In general, this is in the gigahertz range or lower. On the other hand, optical frequencies lie in the hundreds of terahertz, so the assumption that $\Omega/\Omega_0 \ll 1$ is well justified. For gravitational waves this is not necessarily the case. However, we will proceed with the standard quantum optics derivation in the remainder of this section to emphasize the similarity between a quantized gravitational field and a quantized electromagnetic field. In the following subsection, we will see that the essential relation derived from the quantized gravitational field, that between a gravitational wave's strain and the graviton number flux, holds regardless of whether or not we make the standard quantum optics assumption that $\Omega/\Omega_0 \ll 1$ (see \cref{gwFlux,gwFluxBroadband}). 

We now define the modulation frame quadrature operators for the quantized gravitational wave by 
\begin{subequations}
\begin{align}
    \hat{X}(t) := \frac{\hat{a}(t) + \hat{a}^\dagger(t)}{\sqrt{2}} \\
    \hat{P}(t) := \frac{\hat{a}(t) - \hat{a}^\dagger(t)}{i\sqrt{2}},
\end{align}
\end{subequations}
which satisfy the commutation relations
\begin{equation}
\begin{split}
    \left[\hat{X}[\Omega], \hat{P}[\Omega^\prime]\right] = 2\pi i \delta[\Omega - \Omega^\prime]
\end{split}
\end{equation}
with other commutators approximately zero.

The total gravitational wave strain is given by
\begin{equation}\label{quantizedGwStrain}
\begin{split}
    \hat{h}(t) &= \hat{h}^{(+)} + \hat{h}^{(-)} \\
    &= \sqrt{\frac{8 \pi \hbar G}{c^3 \Omega_0 A}} \left( e^{-i \Omega_0 t} \hat{a}(t) + e^{i \Omega_0 t} \hat{a}^\dagger(t) \right) \\
    &= \sqrt{\frac{16 \pi \hbar G}{c^3 \Omega_0 A}} \left( \cos(\Omega_0 t ) \hat{X}(t) + \sin(\Omega_0 t) \hat{P}(t) \right).
\end{split}
\end{equation}
As a result of this expression, we see that $\hat{X}$ and $\hat{P}$ are the two-graviton operators for the cosine and sine quadratures (or amplitude and phase quadratures), by analogy with two-photon quantum optics~\cite{Caves:1985zz,Schumaker:1985zz}.

\subsection{Graviton flux}

In this section, we derive two expressions for the graviton number flux in terms of the power spectral density of gravitational wave strain. The first derivation assumes that the gravitational wave has a carrier frequency $\Omega_0$ and that all frequencies of interest, $\Omega$, lie in a small range around $\Omega_0$ such that $\Omega/\Omega_0 \ll 1$ and \cref{eq:gravitonQuantumOpticsEqs} is valid. This first derivation allows for states such as squeezed states of the gravitational field at the expense of the narrow frequency band assumption. In the second derivation, we instead assume that the different frequency components in the gravitational wave are uncorrelated and lift the narrow-band assumption. The resulting expression then remains valid for states spanning a large frequency range, such as chirped gravitational wave sources and thermal states. Both sets of assumptions and both expressions are valid for the coherent states and Fock states considered in the main text.

\subsubsection{Graviton Flux for Narrow-Band Sources}

The first derivation proceeds as in the analysis of continuous-wave electromagnetic fields in refs. \cite{Blow90,Danilishin2012}.
For a given quantum state of the gravitational wave, we define the mean value and fluctuations of the annihilation operators by 
\begin{equation}
    \hat{a}_\Omega = \bar{a}_\Omega + \delta \hat{a}_\Omega,
\end{equation}
where $\langle \hat{a}_\Omega \rangle = \bar{a}_\Omega$, and $\langle \delta\hat{a}_\Omega \rangle = 0$.

For a coherent state with frequency $\Omega_0$, we have $\bar{a}_\Omega = \bar{a} \delta(\Omega - \Omega_0)$ such that the corresponding mean strain amplitude is (taking $\bar{a}$ to be real)
\begin{equation}
    \langle \hat{h}(t) \rangle = \sqrt{\frac{8\hbar G}{\pi c^3 \Omega_0 A}} \, \bar{a} \cos(\Omega_0 t).
\end{equation}

The graviton number flux operator is given by \cite{Danilishin2012}
\begin{equation}\label{gwFluxDef}
\begin{split}
    \hat{N}_\text{GW}(t) &= \int\limits_0^\infty \frac{\rmd\Omega}{2\pi} \int\limits_0^\infty \frac{\rmd\Omega^\prime}{2\pi} \hat{a}_\Omega^\dagger \hat{a}_{\Omega^\prime} e^{i(\Omega - \Omega^\prime)t} ,
\end{split}
\end{equation}
so the mean number of gravitons in the coherent state is $\bar{a}^2/(2\pi)^2$. We then find that the power radiated by a coherent state of the gravitational field through a cross-sectional area $A$ is $P_\text{GW} = \frac{\hbar \Omega_0 \langle \hat{N}_\text{GW} \rangle}{A} = \frac{c^3 \Omega_0^2}{16 \pi G} \overline{\langle \hat{h}(t) \rangle^2} $, in agreement with the classical result (see \cite[\S{}3.3.2]{Creighton:2011zz}).

In the following sections, it will prove to be convenient to have an expression for the graviton number flux in terms of the power spectral density of the gravitational wave strain. From \cref{gwFluxDef}, we see that
\begin{equation}
    \hat{N}_\text{GW}(t) \approx \hat{a}^\dagger(t) \hat{a}(t).
\end{equation}
For weak-stationary states, we have $\langle \hat{N}_\text{GW}(t) \rangle = \langle \hat{N}_\text{GW}(0) \rangle$ such that
\begin{equation}
\begin{split}
    \langle \hat{N}_\text{GW}(t) \rangle &= \langle \hat{a}^\dagger(0) \hat{a}(0) \rangle \\
    &= \int\limits_{-\infty}^\infty \frac{\rmd\Omega}{2\pi} \int\limits_{-\infty}^\infty \frac{\rmd\Omega^\prime}{2\pi} \langle \hat{a}[\Omega]^\dagger \hat{a}[\Omega^\prime] \rangle \\
    &= \frac{1}{2} \int\limits_{-\infty}^\infty \frac{\rmd\Omega}{2\pi} \int\limits_{-\infty}^\infty \frac{\rmd\Omega^\prime}{2\pi} \Big\langle \left(\hat{X}[\Omega]^\dagger + i \hat{P}[\Omega]^\dagger \right)\\
    & \qquad \qquad \qquad \qquad \qquad \times \left(\hat{X}[\Omega^\prime] + i \hat{P}[\Omega^\prime] \right) \Big\rangle \\
    &= \frac{1}{2} \int\limits_{-\infty}^\infty \frac{\rmd\Omega}{2\pi} \int\limits_{-\infty}^\infty \frac{\rmd\Omega^\prime}{2\pi} \Big\langle \left(\hat{X}[\Omega] + i \hat{P}[\Omega] \right) \\
    & \qquad \qquad \qquad \qquad \qquad \times \left(\hat{X}[\Omega^\prime] + i \hat{P}[\Omega^\prime] \right) \Big\rangle \\
    &= \frac{1}{2} \int\limits_{-\infty}^\infty \frac{\rmd\Omega}{2\pi} \int\limits_{-\infty}^\infty \frac{\rmd\Omega^\prime}{2\pi} \Big( \frac{1}{2} \langle \{ \hat{X}[\Omega], \hat{X}[\Omega^\prime] \} \rangle \\
    & \qquad \qquad \qquad \qquad \quad + \frac{1}{2} \langle \{ \hat{P}[\Omega], \hat{P}[\Omega^\prime] \} \rangle \\
    & \qquad \qquad \qquad \qquad \quad - 2 \pi \delta[\Omega + \Omega^\prime] \Big) \\
    &= \frac{1}{2} \int\limits_{-\infty}^\infty \frac{\rmd\Omega}{2\pi} \left( \bar{S}_{XX}[\Omega] + \bar{S}_{PP}[\Omega] - 1 \right).
\end{split}
\end{equation}
We would now like to relate this expression to the gravitational wave strain amplitude. From \cref{quantizedGwStrain} we have
\begin{equation}
    \hat{h}[\Omega] = \sqrt{\frac{8 \pi \hbar G}{c^3 \Omega_0 A}} \left( \hat{a}[\Omega - \Omega_0] + \hat{a}^\dagger[\Omega + \Omega_0] \right)
\end{equation}
for a weak-stationary GW with fluctuations symmetric about the carrier such that $\bar{S}_{QQ}[\Omega_0 - \Omega] = \bar{S}_{QQ}[\Omega_0 + \Omega]$ with $Q \in \{X,P\}$, we have
\begin{equation}
    \bar{S}_{hh}[\Omega - \Omega_0] = \frac{8 \pi \hbar G}{c^3 \Omega_0 A} \left( \bar{S}_{XX}[\Omega] + \bar{S}_{PP}[\Omega]\right).
\end{equation}
Combining this relation with the expression for graviton flux derived above, we have
\begin{equation}\label{gwFluxWithConstant}
    \langle \hat{N}_\text{GW}(t) \rangle = \int\limits_{-\infty}^\infty \frac{\rmd\Omega}{2\pi} \left( \frac{c^3 \Omega_0 A}{16 \pi \hbar G} \bar{S}_{hh}[\Omega] - \frac{1}{2} \right).
\end{equation}
For a gravitational wave containing more than a few gravitons, we expect the first term in this expression to dominate such that
\begin{equation}
    \label{gwFlux}
    \langle \hat{N}_\text{GW}(t) \rangle = \frac{c^3 \Omega_0 A}{16 \pi \hbar G}\int\limits_{-\infty}^\infty \frac{\rmd\Omega}{2\pi} \bar{S}_{hh}[\Omega].
\end{equation}

\subsubsection{Graviton Flux for Broadband Sources}

In the preceding subsection, we have derived \cref{gwFlux} under the assumption that the relevant gravitational waves occupy a narrow band of frequencies around some carrier. This assumption is valid for certain systems, such as the kinds of binary white dwarf systems expected to be observed by LISA, but not for rapidly evolving sources such as the binary black hole systems observed by LIGO. In this subsection, we will lift the narrow-band assumption and derive a formula similar to \cref{gwFlux}. Instead, we assume that there are no correlations between the gravitational waves' different frequency components such that $\langle \hat{a}_\Omega \hat{a}_{\Omega^\prime} \rangle = 0$ and $\langle \hat{a}_\Omega^\dagger \hat{a}_{\Omega^\prime} \rangle = 0$ unless $\Omega = \Omega^\prime$.

As in the previous subsection, we begin with \cref{gwFluxDef}. We also define $H_\omega = \sqrt{8\pi\hbar G / (c^3 |\omega| A)}$ for notational simplicity. Using the assumption of no correlations between distinct frequency components, we have 

\begin{widetext}
\begin{equation}
\begin{split}
    \langle \hat{N}_\text{GW}(t) \rangle &= \int\limits_0^\infty \frac{\rmd\Omega}{2\pi} \int\limits_0^\infty \frac{\rmd\Omega^\prime}{2\pi} \langle \hat{a}_\Omega^\dagger \hat{a}_{\Omega^\prime} \rangle e^{i(\Omega - \Omega^\prime)t} \\
    &= \int\limits_0^\infty \frac{\rmd\Omega}{2\pi} \int\limits_0^\infty \frac{\rmd\Omega^\prime}{2\pi} \langle \hat{a}_\Omega^\dagger \hat{a}_{\Omega^\prime} \rangle \\
    &= \frac{1}{2} \int\limits_0^\infty \frac{\rmd\Omega}{2\pi} \int\limits_0^\infty \frac{\rmd\Omega^\prime}{2\pi} \left( \langle \hat{a}_\Omega^\dagger \hat{a}_{\Omega^\prime} \rangle + \langle \hat{a}_{\Omega^\prime} \hat{a}_\Omega^\dagger \rangle - 2\pi \delta(\Omega - \Omega^\prime)  \right)\\
    &= \frac{1}{2} \int\limits_{-\infty}^\infty \frac{\rmd \omega}{2\pi} \int\limits_0^\infty \frac{\rmd\Omega}{2\pi} \int\limits_0^\infty \frac{\rmd\Omega^\prime}{2\pi} 2\pi \left( \delta(\omega+\Omega) \frac{H^2_{\Omega}}{H^2_{\omega}} \langle \hat{a}_\Omega^\dagger \hat{a}_{\Omega^\prime} \rangle + \delta(\omega -\Omega) \frac{H^2_{\Omega}}{H^2_{\omega}} \langle \hat{a}_{\Omega^\prime} \hat{a}_\Omega^\dagger \rangle  \right) - \int\limits_0^\infty \frac{\rmd\Omega}{2\pi} \\
    &= \frac{1}{2} \int\limits_{-\infty}^\infty \frac{\rmd \omega}{2\pi} \int\limits_{-\infty}^\infty \rmd t \, e^{i\omega t} \frac{1}{H^2_{\omega}} \int\limits_0^\infty \frac{\rmd\Omega}{2\pi} \int\limits_0^\infty \frac{\rmd\Omega^\prime}{2\pi} \left\langle H_{\Omega} \left( \hat{a}_\Omega e^{-i\Omega t} + \hat{a}_\Omega^\dagger e^{i \Omega t} \right) H_{\Omega^\prime}\left( \hat{a}_{\Omega^\prime} + \hat{a}_{\Omega^\prime}^\dagger \right) \right\rangle - \int\limits_0^\infty \frac{\rmd\Omega}{2\pi} \\
    &= \frac{1}{2} \int\limits_{-\infty}^\infty \frac{\rmd \omega}{2\pi} \frac{1}{H^2_{\omega}} \int\limits_{-\infty}^\infty \rmd t \, e^{i\omega t} \left\langle \left\{ \hat{h}(t), \hat{h}(0) \right\} \right\rangle - \int\limits_0^\infty \frac{\rmd\Omega}{2\pi} \\
    &= \frac{1}{2} \int\limits_{-\infty}^\infty \frac{\rmd \omega}{2\pi} \left( \frac{\bar{S}_{hh}[\omega]}{H^2_{\omega}}  - 1 \right) \\
    &= \int\limits_{-\infty}^\infty \frac{\rmd \Omega}{2\pi} \left( \frac{c^3 \Omega A}{16 \pi \hbar G} \bar{S}_{hh}[\Omega]  - \frac{1}{2} \right).
\end{split}
\end{equation}
\end{widetext}
We observe that this final expression is the same as \cref{gwFluxWithConstant} with $\Omega_0 \rightarrow \Omega$. Additionally, if the gravitational wave has a large number of gravitons, we can neglect the constant term and have 
\begin{equation}\label{gwFluxBroadband}
    \langle \hat{N}_\text{GW} \rangle = \int_{-\infty}^\infty \frac{\rmd \Omega}{2\pi} \frac{c^3 \Omega A}{16 \pi \hbar G} \bar{S}_{hh}[\Omega],
\end{equation}
which is similar to \cref{gwFlux} with $\Omega_0 \rightarrow \Omega$.

We emphasize again that \cref{gwFluxBroadband} assumes that different frequency components are uncorrelated instead of assuming that the strain spectrum is narrow-band.

\subsubsection{Graviton Flux from Astrophysical Sources}

To determine the total number of gravitons in a gravitational wave, we need to determine the relevant area factor, $A$.
So far, we have been considering plane waves, but astrophysical gravitational wave sources generally emit spherical waves.
For a binary system of reduced mass $\mu$ in the $xy$ plane executing a circular orbit with frequency $\Omega_0/2$ and radius $a$, the gravitational waves at a distance $R$ and inclination $\theta$ are, in the post-Newtonian approximation~\cite[\S{}3]{Maggiore07},
\begin{subequations}
\begin{align}
    h_+(t) &= \frac{G \mu \Omega_0^2 a^2}{c^4 R} \left( \frac{1 + \cos^2{\theta}}{2} \right) \cos(\Omega_0 t) \\
    h_\times(t) &= \frac{G \mu \Omega_0^2 a^2}{c^4 R} \cos{\theta} \sin(\Omega_0 t)
\end{align}
\end{subequations}
and the system radiates a cycle-averaged intensity
\begin{equation}
\begin{split}
    \frac{\rmd{} P_{\text{quad}}}{\rmd A} &= \frac{c^3}{16\pi G} \left\langle \dot{h}_+^2 + \dot{h}_\times^2 \right\rangle \\
    &= \frac{G^2 \mu^2 \Omega_0^6 a^4}{32 \pi c^5 R^2} \left[ \left(\frac{1 + \cos^2{\theta}}{2}\right)^2 + \cos^2{\theta}\right]
\end{split}
\end{equation}
with total power
\begin{equation}
    P_{\text{quad}} = \int\limits_0^{2\pi} \rmd\phi \int\limits_0^\pi \rmd\theta \, \sin{\theta} \, R^2 \frac{\rmd{}P_{\text{quad}}}{\rmd{}A} = \frac{G^2 \mu^2 \Omega_0^6 a^4}{10 c^5}.
\end{equation}
Conversely, the cycle-averaged intensity that can be absorbed by a detector with antenna tensor $D_{ij}$ (\cref{hmunu}), and located at inclination $\theta_0$ relative to the source, is
\begin{equation}
\begin{split}
    \frac{\rmd{} P_{\text{plane}}}{\rmd A} &= \frac{c^3}{16\pi G} \left\langle \left(D_+ \dot{h}_+\right)^2 + \left(D_\times \dot{h}_\times\right)^2 \right\rangle \\
    &= \frac{G^2 \mu^2 \Omega_0^6 a^4}{32 \pi c^5 R^2} \underbrace{\left[ D_+^2 \left(\frac{1 + \cos^2{\theta_0}}{2}\right)^2 + D_\times^2 \cos^2{\theta_0}\right]}_{\equiv F},
\end{split}
\end{equation}
where $D_+$ and $D_\times$ are the two independent components of the tensor, and $F$ is evidently an order-unity factor determined by the detector geometry and its orientation relative to the source.
The relevant area for \cref{gwFlux} can then be found by
\begin{equation}
    A = \frac{P_{\text{quad}}}{\rmd{}P_{\text{plane}} / \rmd{}A} = \frac{16\pi R^2}{5F}.
\end{equation}
Hence, we can relate the detector's strain spectral density $\bar{S}_{hh}$ to the graviton flux emitted by the source:
\begin{equation}
    \label{eq:Ngw}
    \langle \hat{N}_\text{GW}(t) \rangle = \frac{c^3 R^2 \Omega_0}{5 \hbar G F}\int\limits_{-\infty}^\infty \frac{\rmd\Omega}{2\pi} \bar{S}_{hh}[\Omega].
\end{equation}
Since the tensorial strain $h_{\mu\nu}(t)$ generically scales as $1/R$ for $R \gg a$, and since $\bar{S}_{hh}$ scales with the square of the strain, \cref{eq:Ngw} is overall independent of $R$, as expected.

\section{Graviton flux using the Poynting vector for gravitational radiation}

In this section we derive for completeness a classical gauge-invariant description of the energy flux of the gravitational radiation that is quantized to form the graviton number flux operator in \cref{quantderivation}.

\subsection{Energy content to first order in the metric perturbation}
\label{firstorder}
We take as our starting point, as in \cref{classict}, the Einstein--Hilbert action of pure gravity (neglecting any scalar or matter fields that may form an energy--momentum tensor). If we now expand the Ricci tensor up to first order in the metric perturbation, we obtain
\begin{equation}
    R_{\mu \nu}[g] = R^{(1)}_{\mu \nu}[h],
\end{equation}
where to first order in the perturbation (simplifying both in the TT gauge and the harmonic gauge), we have
\begin{equation}
    R^{(1)}_{\mu \nu}[h] = -\frac{1}{2}\partial^\rho\partial_\rho h_{\mu \nu}.
\end{equation}
Therefore, if we use this approximation in the Einstein--Hilbert action of \cref{seh}, and maintaining terms up to $\mathcal{O}(h^2)$, we obtain:
\begin{equation}\label{seh11}
\begin{split}
    \mathcal{S}^{(1)}_{\mathrm{EH}}[g] &= \frac{c^4}{32 \pi G} \int\limits_M \mathrm{~d}^4 x (\eta^{\mu \nu} + h^{\mu \nu})  R_{\mu \nu}^{(1)}[h],\\
    &= \frac{c^4}{32 \pi G} \int\limits_M \mathrm{~d}^4 x (\eta^{\mu \nu} + h^{\mu \nu})  \partial^\rho \partial_\rho h_{\mu \nu}.
\end{split}
\end{equation}
If we now further simplify for the gauge-fixing conditions, and extremize the action, we obtain (up to surface terms):
\begin{equation}
    \delta \mathcal{S}^{(1)}_{\mathrm{EH}}[g] = -\frac{c^4}{16 \pi G} \int\limits_M \mathrm{~d}^4 x (\delta h_{\mu \nu})  \partial^\rho \partial_\rho h_{\mu \nu}.
\end{equation}
Here we have used the identity $\delta g^{a b}=-g^{a c} g^{b d} \delta g_{c d}$, to lowest order in the perturbation is $\delta h^{\mu \nu }=-\eta^{\rho \mu}\eta^{\nu \beta}  \delta h_{\rho \beta}$. Therefore, extremizing the action by imposing 
\begin{equation}
    \frac{\delta \mathcal{S}^{(1)}_{\mathrm{EH}}[g]}{\delta h_{\mu \nu}} = 0
\end{equation}
directly implies the source-free wave equation for gravitational radiation:
\begin{equation}
    \partial^\rho \partial_\rho h_{\mu \nu} = 0.
\end{equation}
However, there is no energy--momentum tensor involved, in particular because our starting point in \cref{seh11}, was the action for pure gravity with no scalar fields that would model any matter field distribution. In order to understand the energy content of gravitational radiation, we can either use the approach currently in \cref{quant1}, in which a Hamiltonian density is constructed from the canonical pairs, or we can examine the Ricci tensor up to second order, and use the second order terms as an effective energy momentum contribution to the Einstein--Hilbert action. 

\subsection{Energy content to second order in the metric perturbation}
We now examine second order corrections to the Ricci tensor
\begin{equation}
    R^{(2)}_{\mu \nu}[g] = R_{\mu \nu}^{(1)}[h] +  R_{\mu \nu}^{(1)}[h^{(2)}] + R_{\mu \nu}^{(2)}[h].
\end{equation}
Here, $R_{\mu \nu}^{(2)}[h]$ corresponds to the terms that are quadratic in the first order perturbation, while $R_{\mu \nu}^{(1)}[h^{(2)}]$ corresponds to the terms that are linear in the second order perturbation. If we now use this approximation in the Einstein--Hilbert action of \cref{seh}, and maintain terms up to $\mathcal{O}\left(h^3\right)$, we obtain
\begin{equation}\label{seh22}
\begin{split}
    \mathcal{S}_{\mathrm{EH}}[g] &= -\frac{c^4}{16 \pi G} \int\limits_M \rmd{}^4 x \left(\eta^{\mu \nu} + h^{\mu \nu}\right)  R_{\mu \nu}^{(2)}[g] \\
    &= -\frac{c^4}{16 \pi G} \int\limits_M \rmd{}^4 x \bigg(h^{\mu \nu} R_{\mu \nu}^{(1)}[h] \\
    & \qquad \qquad +\left(\eta^{\mu \nu} + h^{\mu \nu}\right)\left(R_{\mu \nu}^{(1)}[h^{(2)}] + R_{\mu \nu}^{(2)}[h]\right)\bigg) \\
    &= \mathcal{S}^{(1)}_{\mathrm{EH}}[h]  + \mathcal{S}^{(2)}_{\mathrm{EH}}[h],
\end{split}
\end{equation}
where in the second line we have applied the TT-gauge condition ${h^\mu}_\mu = 0$, and we have also defined
\begin{subequations}
\label{seh2}
\begin{align}
    \mathcal{S}^{(1)}_{\mathrm{EH}}[h] &= -\frac{c^4}{16 \pi G} \int\limits_M \rmd{}^4 x \; h^{\mu \nu}  \left(R_{\mu \nu}^{(1)}[h]\right) \\
    \mathcal{S}^{(2)}_{\mathrm{EH}}[h] &= -\frac{c^4}{16 \pi G} \int\limits_M \rmd{}^4 x \left(\eta^{\mu \nu}+h^{\mu \nu}\right) \\ & \qquad \qquad \qquad \qquad \times\left(R_{\mu \nu}^{(1)}\left[h^{(2)}\right]+R_{\mu \nu}^{(2)}[h]\right).
\end{align}
\end{subequations}
Note that we have (in the TT and harmonic gauge),
\begin{subequations}
\begin{align}
    R_{\mu \nu}^{(1)}[h] &= -\frac{1}{2} \partial^\rho \partial_\rho h_{\mu \nu}  \\
    R_{\mu \nu}^{(2)}[h] & =\frac{1}{2} h^{\rho \sigma} \partial_\mu \partial_\nu h_{\rho \sigma}-h^{\rho \sigma} \partial_\rho \partial_{(\mu} h_{\nu) \sigma} \\
    & \quad \, +\frac{1}{4} \partial_\mu h_{\rho \sigma} \partial_\nu h^{\rho \sigma}+\partial^\sigma h^\rho{ }_\nu \partial_{[\sigma} h_{\rho] \mu}.
\end{align}
\end{subequations}
Now, imposing that the Einstein equations are satisfied to first order by applying the results from \cref{firstorder}, the action simplifies to $\mathcal{S}_{\text{EH}}[g] = \mathcal{S}^{(2)}_{\text{EH}}[h]$.
Typically, the energy--momentum tensor for matter is identified from the action for matter through
\begin{align}
    \mathcal{S}_{\mathrm{matter}} &= \int\limits_M \rmd{}^4 x \,\sqrt{-g} \, \mathcal{L}_{\mathrm{matter}} \\
    T^{\mu\nu} &= \frac{2}{\sqrt{-g}} \frac{\delta \mathcal{S}_{\text{matter}}}{\delta g_{\mu\nu}}\\
    \delta \mathcal{S}_{\text{matter}} &= \frac{1}{2} \int\limits_M \rmd{}^4 x \,\sqrt{-g} \, T^{\mu\nu} \,\delta g_{\mu\nu},
\end{align}
where the Einstein equations are then derived via $\frac{\delta}{\delta g_{\mu\nu}}\left(\mathcal{S}_{\text{EH}}+\mathcal{S}_{\text{matter}}\right) = 0$~\cite[{\S}VI]{Zee:2013dea}. Therefore, if we now consider the variation of this action up to second order, we obtain
\begin{equation}
\begin{split}
    &\delta \mathcal{S}^{(2)}_{\mathrm{EH}}[h] = \\
    & \quad -\frac{c^4}{16 \pi G} \int\limits_M \rmd{}^4 x\bigg(\left(\eta^{\mu \nu}+\delta h^{\mu \nu}\right)\left(R_{\mu \nu}^{(1)}[h^{(2)}]+R_{\mu \nu}^{(2)}[h]\right) \\
    & \qquad \qquad \qquad \qquad \quad +  h^{\mu \nu}\delta R_{\mu \nu}^{(2)}[h] \bigg).
\end{split}
\end{equation}
From this, we can identify the the part that contains the effective form of the gravitational field of the metric perturbation to second order:
\begin{equation}
    \frac{\delta \mathcal{S}^{(2)}_{\mathrm{EH}}[h]}{\delta h_{\mu \nu}} = R_{\mu \nu}^{(1)}[h^{(2)}] \equiv G_{\mu \nu}^{(2)},
\end{equation}
which we identify as the effective second-order Einstein tensor, for the second order perturbation of the linearized gravitational field. (This expression is valid assuming the second-order tensor is also traceless). We therefore make the association 
\begin{equation}
    \delta \mathcal{S}^{(2)}_{\mathrm{matter}}[h] = -\frac{c^4}{16 \pi G} \int \rmd{}^4 x \left(\delta h^{\mu \nu} R_{\mu \nu}^{(2)}[h] + h^{\mu \nu} \delta R_{\mu \nu}^{(2)}[h] \right).
\end{equation}
This yields an effective energy--momentum tensor
\begin{equation}\label{tuv1}
    t_{\mu \nu} = \frac{\delta \mathcal{S}^{(2)}_{\mathrm{matter}}[h]}{\delta h_{\mu \nu}} = -\frac{c^4}{8\pi G } (R_{\mu \nu}^{(2)}[h] ),
\end{equation}
where the second term involving $\delta R_{\mu \nu}^{(2)}[h]$ is eliminated after applying $\frac{\delta \mathcal{S}_{\mathrm{EH}}^{(2)}[h]}{\delta h_{\mu \nu}}$, and simplifying the resulting contractions after applying the trace-free ${h^\mu}_{\mu} = 0$, and harmonic gauge $\partial^\nu h_{\nu \mu} = 0$ conditions. There is one further subtlety, in that the effective energy--momentum tensor is not actually gauge invariant. Although \( t_{\mu\nu} \) is not gauge invariant on its own, it can be rendered gauge invariant through an averaging procedure. As in Ref.~\cite{Reall13}, consider an open region \( \mathcal{R} \subset \mathbb{R}^4 \) containing the origin, with co-ordinate size \( a \) in all directions. We take \( W(y) \) to be be a smooth weight function on \( \mathbb{R}^4 \) satisfying the following properties: \( W = 0 \) outside \( \mathcal{R} \), \( W > 0 \) inside \( \mathcal{R} \), and
\begin{equation}
    \int\limits_\mathcal{R} \rmd{}^4y \; W(y) = 1.
\end{equation}
The average of a tensor field \( X_{\mu\nu} \) at the point \( x \) is then defined as
\begin{equation}
    \langle X_{\mu\nu} \rangle(x) = \int\limits_\mathcal{R} \rmd{}^4 y \; W(x - y) X_{\mu\nu}(y).
\end{equation}
This averaging is performed in a region far from sources, where gravitational radiation is present with a characteristic wavelength \( \lambda \), and we assume \( \lambda \ll a \). Suppose the typical magnitude of the tensor components \( X_{\mu\nu} \) is \( X \). Since the radiation has wavelength \( \lambda \), the derivatives \( \partial_\alpha X_{\mu\nu} \) are expected to scale as \( X / \lambda \). Now consider the average of the derivative \( \partial_\alpha X_{\mu\nu} \):
\begin{equation}
    \langle \partial_\alpha X_{\mu\nu} \rangle(x) = \int\limits_\mathcal{R} (\partial_\alpha W)(x - y) X_{\mu\nu}(y) \, d^4y,
\end{equation}
where we have integrated by parts, making use of the fact that \( W = 0 \) on the boundary \( \partial \mathcal{R} \). The derivative \( \partial_\alpha W \) typically has magnitude \( W / a \), so the integrand is of order \( X / a \). Therefore, choosing \( a \gg \lambda \) ensures that the averaging procedure suppresses derivatives by a factor \( \lambda / a \ll 1 \). As a result, total derivatives can be neglected under the averaging integral. This allows us to perform integration by parts inside the averaging operation:
\begin{equation}
\langle A \, \partial_\mu B \rangle = \langle \partial_\mu (A B) \rangle - \langle (\partial_\mu A) B \rangle 
\approx - \langle (\partial_\mu A) B \rangle,
\end{equation}
where the total derivative term has been dropped due to its suppression in the averaging process. If we apply this averaging procedure to \cref{tuv1}, and again imposing the harmonic gauge and TT gauge conditions, we obtain:
\begin{equation}
    \left\langle t_{\mu \nu}\right\rangle=\frac{c^4}{32 \pi G}\left\langle\partial_\mu h_{\rho \sigma} \partial_\nu h^{\rho \sigma}\right  \rangle.
\end{equation}
From this, we can identify the energy density as the $t_{00}$ component, and the Poynting vector as the $t_{0i}$ component. This would give the following power per unit area, for a 1-D plane wave $h_{i j}(t)=h_{i j}^{(0)} \cos [\Omega_0(z / c-t)]$ (and averaging over several wavelengths):
\begin{equation}
\left\langle t_{0 i}\right\rangle_\mathrm{avg} = \frac{c^3\Omega_0^2}{64 \pi G},
\end{equation}
which re-produces the classical average power flux (and $\mathrm{avg}$ indicates averaged over several wavelengths of the plane wave). We hence-forth drop the averaging symbols for readability, leaving the average implicit.  We therefore, can write the Poynting vector in canonical form as (using $\Pi_{ij} = \frac{c^2}{32 \pi G} \dot{h}_{i j}$):
\begin{equation}
    t_{0 i} = c^2 \Pi_{\rho \sigma} \partial_i h^{\rho \sigma}.
\end{equation}
This expression serves as the gravitational-wave analogue of the canonical description of the Poynting vector. However, unlike in electromagnetism, it cannot be simply written as a cross product of the canonical coordinates. Importantly, it yields a gauge-invariant form of the energy carried by propagating gravitational radiation. When the quantization procedure from Section~\ref{quant1} is applied, this expression leads to the graviton flux number operator. Additionally, it provides a gauge-invariant energy density suitable for applying the energy conservation argument from Refs.~\cite{TobPik24,tobberg}, for a graviton counter to provide evidence of quantization.

\section{Interaction between a bar and gravitational radiation in TT gauge}\label{interacthamil}

Here, we derive the interaction between a resonant mass detector and quantized gravitational radiation for completeness. We express the interaction in the local inertial frame in terms of the linearized gravitational field operator in the TT gauge. This allows the previously derived quantization of the linearized gravitational field in the TT gauge to be applied to describe the interaction between resonant mass detectors and quantized gravitational radiation. 

We model the bar as an elastic medium with Lam\'e elasticity parameters $\lambda$ and $\mu$ described by an effective field theory for elastic media with an elastic displacement field $\boldsymbol{u}(x, y, z, t)$. We assume the Lam\'e parameters are constants, that the spatial extent of the bar is much less than one wavelength of the gravitational field, and work to leading order in the displacement field and $v_\text{s}/c$, with $v_\text{s}$ the speed of sound in the bar. The bare Lagrangian density of the bar is then \cite{belgacem2024coupling}
\begin{equation}
\begin{split}
    \mathcal{L}^\text{bar} = \frac{1}{2} \Big(& \rho \dot{u}_i \dot{u}^i - \lambda (\partial_i u^i)^2  \\
    &- \frac{1}{2} \mu (\partial_i u_j +\partial_j u_i) (\partial^i u^j +\partial^j u^i) \Big),
\end{split}
\end{equation}
and the interaction Lagrangian between the bar and a gravitational wave in the TT gauge is 
\begin{equation}
    \mathcal{L}^\text{int} = \frac{1}{2}h^\text{TT}_{ij}T^{ij},
\end{equation}
where the components of the stress-energy tensor $T_{ij}$ are given by 
\begin{equation}
    T_{ij}=-\lambda\delta_{ij}\partial_k u^k-\mu (\partial_i u_j + \partial_j u_i).
\end{equation}
From the total Lagrangian $\mathcal{L} = \mathcal{L}^\text{bar} + \mathcal{L}^\text{gw} + \mathcal{L}^\text{int}$, the bar's equation of motion is given by 
\begin{equation}\label{eq:barEftEom}
    \rho \ddot{u}_i = \lambda \partial_i \partial_j u^j + \mu \partial_j(\partial_i u^j + \partial^j u_i),
\end{equation}
with the boundary condition
\begin{equation}\label{eq:barEftBc}
    n^j \left( \lambda \delta_{ij} \partial_k u^k + \mu (\partial_i u_j + \partial_j u_i + h_{ij}^\text{TT}) \right) = 0,
\end{equation}
where $\mathbf{n}$ is the unit vector normal to the bar's surface. From \cref{eq:barEftEom,eq:barEftBc}, we see that in the TT-gauge the gravitational wave strain does not enter into the equations of motion for the bar directly, but instead couples to its boundary condition where there is an inhomogeneity in the elastic medium. 

However, it is convenient to include the gravitational wave strain in the bar's equations of motion rather than its boundary condition. To do so, we define the lab-frame coordinates by
\begin{equation}
    x_i^\text{lab} = x_i + \frac{1}{2} h_{ij}(t)x^j
\end{equation}
and the modified strain density (in the lab coordinate frame) $\delta \boldsymbol{u}$ by
\begin{equation}
    \delta u_i := u_i + \frac{1}{2} h_{ij}^\text{TT}(t)x^j.
\end{equation}
In both frames, the strain satisfies the transverse-traceless conditions.
In the lab frame and in terms of $\delta \boldsymbol{u}$, the bar's equation of motion is \cite{belgacem2024coupling,hudelist2023relativistic}
\begin{equation}\label{eq:barEomLab}
    \rho \delta \ddot{u}_i = \lambda \partial_i \partial_j \delta u^j + \mu \partial_j(\partial_i \delta u^j + \partial^j \delta u_i) + \frac{\rho}{2} \ddot{h}_{ij} x^j
\end{equation}
with the boundary condition
\begin{equation}\label{eq:barBcLab}
    n^j \left( \lambda \delta_{ij} \partial_k \delta u^k + \mu (\partial_i \delta u_j + \partial_j \delta u_i) \right) = 0.
\end{equation}

We now expand the displacement field into a series of eigenmodes as done in Ref.~\cite{HiraFuji76}:
\begin{equation}
    \delta\boldsymbol{u}_x(x, y, z, t)=\sum_n b_n(t) \boldsymbol{w}_n(x, y, z).
\end{equation}
We now restrict to the simple case of one-dimensional acoustic oscillations and consider sinusoidal displacement fields:
\begin{equation}
    \delta u_x(t, x)=\sum_{n}  b_{n }(t) \sin \left[\frac{\pi x}{L}(2 n+1)\right];
\end{equation}
before substitution, we set $\mu$ = 0, and identify the speed of sound in the material as $v_{\text{s}} = \sqrt{\lambda/\rho}$.
It can be verified that these modes satisfy the boundary condition \cref{eq:barBcLab} for a one dimensional bar.
For one-dimensional oscillations (and only considering the coupling to the $xx$ component of the gravitational wave for simplicity), the equation of motion \cref{eq:barEomLab} becomes
\begin{equation}
\delta \ddot{u}_{x} - v_s^2  \frac{\partial^2}{\partial x^2} \delta u_{x} = \frac{1}{2}   \ddot{h}_{xx} x.
\end{equation}
We now substitute in the mode expansion, and then using the orthogonality relation 
\begin{equation}
    \int\limits_{-L / 2}^{L / 2} \!\rmd{}x \; \sin{\left[\frac{\pi x}{L}(2 n+1)\right]} \sin{\left[\frac{\pi x}{L}(2 m+1)\right]} = \frac{L}{2} \delta_{n,m},
\end{equation}
the equation of motion simplifies to the following ODE for the normal modes $b_n(t)$:
\begin{equation} \label{eom111}
\ddot{b}_{n}(t)+b_{n }(t)\left[v_{\text{s}}^2\left(\frac{\pi(2 n+1)}{L}\right)^2\right]=\frac{(-1)^n}{(2 n+1)^2}\left(\frac{2 L}{\pi^2}\right) \ddot{h}_{x x},
\end{equation}
where we can identify 
\begin{equation}
    \omega_n = v_{\text{s}} \left(\frac{\pi(2 n+1)}{L}\right)
\end{equation}
as the frequency of the normal mode. This is simply the equation of a simple harmonic oscillator driven by the force 
\begin{equation}
    F = M\frac{(-1)^n}{(2 n+1)^2}\left(\frac{2 L}{\pi^2}\right) \ddot{h}_{x x},
\end{equation}
The above force can now be used to find the Hamiltonian via $H = -\partial F /\partial x$, which results in the following interaction Hamiltonian (as considered by Maggiore~\cite{MaggioreMichele2007GwV1}):
\begin{equation} \label{hintqg}
    H_{\mathrm{int}} = \frac{(-1)^n}{(2 n+1)^2}\left(\frac{2ML}{\pi^2}\right) \ddot{h}_{x x}{b}_{n}.
\end{equation}
We can now proceed to quantize the bar's acoustic mode $b_{n}$, and the gravitational wave with strain field $\ddot{h}_{x x}$, and then derive the Heisenberg equations of motion, for which we find the following equation for the bar's quantized normal mode:
\begin{equation}
    \ddot{\hat{b}}_n + \omega_n^2\hat{b}_n = \frac{(-1)^n}{(2 n+1)^2}\left(\frac{2L}{\pi^2}\right) \ddot{\hat{h}}_{x x}. 
\end{equation}
However, we have neglected the dissipation terms. We see that the equations of motion for $\hat{b}_n$  correspond to a quantized form of the classical equation of motion in \cref{eom111}. Hence-forth we will restrict to the fundamental mode of oscillation and re-define $\hat{b}_1 = \hat{x}$, with frequency of oscillation $\omega_m$, yielding the equation of motion for the fundamental mode:
\begin{equation}
    \ddot{\hat{x}} + \omega_m^2\hat{x} = \left(\frac{2L}{\pi^2}\right) \ddot{\hat{h}}_{x x}.
\end{equation}

\section{Optical cavities as gravitational-wave antennae}\label{app:cavityGwAntenna}

In this section we consider how a laser interferometer operated as a GW antenna responds to an incident GW field, either in the case of homodyne read out or of photon counting.
For our purposes, the physics of such an antenna is sufficiently captured by modeling it as a simple Fabry--Perot cavity with a single movable test mass.

\subsection{Equations of motion}

The coupled, time-domain equations of motion for a single Fabry--Perot cavity operating as a GW antenna, derived from the interaction Hamiltonian with quantized gravitational radiation in the TT gauge are~\cite{PangChen18}
\begin{subequations}
\label{eq:antennaTimeDomain}
\begin{align}
    \partial_t \hat{\alpha}_1 &= - \Delta \hat{\alpha}_2 - \kappa \hat{\alpha}_1 + \sqrt{2 \kappa} \hat{\alpha}_1^\text{in} \\
    \partial_t \hat{\alpha}_2 &= \Delta \hat{\alpha}_1 - \kappa \hat{\alpha}_2 + \sqrt{2 \kappa} \hat{\alpha}_2^\text{in} - \frac{\omega_0 \bar{\alpha}}{2} \hat{h} - \frac{\omega_0 \bar{\alpha}}{L} \hat{q} \\
    \partial_t \hat{q} &= \frac{1}{m} \hat{p} - \gamma_{\text{m}} \hat{q} + \sqrt{\frac{\hbar \gamma_{\text{m}}}{m \omega_{\text{m}}}} \hat{\tilde{q}}_\text{in} \\
    \partial_t \hat{p} &= -m \omega_{\text{m}}^2 \hat{q} - \gamma_{\text{m}} \hat{p} + \frac{\hbar \omega_0 \bar{\alpha}}{L} \hat{\alpha}_1 + \sqrt{\hbar \gamma_{\text{m}} m \omega_{\text{m}}} \, \hat{\tilde{p}}_\text{in},
\end{align}
\end{subequations}
where $\hat{\alpha}_1= \frac{1}{\sqrt{2}}\left(\hat{d}+\hat{d}^{\dagger}\right)$ and $\hat{\alpha}_2 =-\frac{i}{\sqrt{2}}\left(\hat{d}-\hat{d}^{\dagger}\right)$ are the linearized circulating amplitude and phase quadratures of the optical field with creation and annihilation operators $\hat{d}$ and $\hat{d}^{\dagger}$ in a frame displaced by the mean field amplitude $\bar{\alpha}$.\footnote{Compared to Ref.~\cite{PangChen18}, our normalization of $\hat{\alpha}_{1,2}$ differs by a factor of $\sqrt{\hbar}$, and hence the Hamiltonian generating the optomechanical coupling for us reads $\hat{H}_{\text{OM}} = -\hbar\omega_0 [(\bar{\alpha}+\hat{\alpha}_1)^2 + \hat{\alpha}_2^2)]\hat{q}/2L$.
The relation of $\bar{\alpha}$ to intracavity power $P_{\text{cav}}$ can by found by noting that the radiation force on the (reflective) test mass is $2 P_\text{cav} /c$; thus $\bar{\alpha}^2 = (4L/c) P_{\text{cav}}/\hbar\omega_0$.\label{fn:normalization}}
Here, $\Delta$ is the optical detuning of the cavity, $\kappa$ is its decay rate, its resonant frequency is $\omega_0$, and its length is $L$.
$\hat{q}$ and $\hat{p}$ are the position and momentum of the test mass with mass $m$, with mechanical resonance frequency $\omega_{\text{m}}$ and mechanical damping rate $\gamma$.
Using the interaction Hamiltonian derived in the TT gauge from Ref.~\cite{PangChen18}, allows the quantization of the linearized gravitational field in the TT gauge to be directly applied from \cref{quantderivation}.  $\hat{\tilde{q}}_\text{in}$ and $\hat{\tilde{p}}_\text{in}$ are the in-coupled mechanical noise modes due to the test mass damping, which have units of $\sqrt{\text{phonons}/\text{second}}$, in contrast with $\hat{q}$ and $\hat{p}$, which have units of displacement and momentum, respectively.

The cavity's output fields are related to its input and circulating fields by
\begin{equation}\label{eq:inOut}
    \hat{\alpha}_j^\text{out} = \hat{\alpha}_j^\text{in} - \sqrt{2 \kappa} \hat{\alpha}_j,
\end{equation}
where $j \in \{1,2\}$.
Taking the Fourier transform of \cref{eq:antennaTimeDomain} and using \cref{eq:inOut}, we arrive at the frequency-domain evolution of the output optical quadratures. For simplicity, we assume that the antenna is operated on-resonance so $\Delta = 0$, as in LIGO. In this case, the detector's output quadratures are given by
\label{eq:gwAntennaOutputQuads}
\begin{equation}
\begin{split}
    \hat{\alpha}_1^\text{out}[\Omega] =& -e^{2 i \beta [\Omega]} \hat{\alpha}_1^\text{in}[\Omega] \\
    \hat{\alpha}_2^\text{out}[\Omega] =&-e^{2 i \beta [\Omega]}\left(\hat{\alpha}_2^\text{in}[\Omega] - e^{2 i \Xi [\Omega]}\mathcal{K}[\Omega] \mathcal{X}[\Omega] \hat{\alpha}_1^\text{in}[\Omega] \right) \\
    &+ e^{i \beta[\Omega]} \sqrt{2 \mathcal{K}[\Omega]} \frac{\hat{h}[\Omega]}{h_\text{SQL}[\Omega]} \nonumber \\
    &+ \sqrt{\frac{2 \hbar \gamma_{\text{m}} \kappa}{m}} \frac{\bar{\alpha} \omega_0}{L (\kappa - i \Omega) \left( (\gamma_{\text{m}} - i \Omega)^2 + \omega_{\text{m}}^2 \right)} \\
    & \qquad \qquad \times \left( \frac{\gamma_{\text{m}} - i \Omega}{\sqrt{\omega_{\text{m}}}} \hat{\tilde{q}}_\text{in}[\Omega] + \sqrt{\omega_{\text{m}}} \hat{\tilde{p}}_\text{in}[\Omega] \right),
\end{split}
\end{equation}
where $\beta$ is the standard frequency-dependent phase shift acquired by light interacting with an optical cavity near resonance given by~\cite{Kimble2001}
\begin{equation}
    \beta[\Omega] \equiv \arctan(\Omega/\kappa);
\end{equation}
$\mathcal{K}$ is the cavity-enhanced optomechanical coupling, given by
\begin{equation}
    \mathcal{K}[\Omega] \equiv \frac{2 \hbar \kappa \bar{\alpha}^2 \omega_0^2}{L^2 m \Omega^2 \left(\kappa^2 + \Omega^2 \right)};
\end{equation}
$h_\text{SQL}$ is the ``standard quantum limit'' strain amplitude, given by
\begin{equation}
    h_\text{SQL}[\Omega] \equiv \sqrt{\frac{8 \hbar}{m \Omega^2 L^2}};
\end{equation}
$\mathcal{X}$ is the modification to the optomechanical coupling due to a finite mechanical resonance frequency and damping:
\begin{equation}
\begin{split}
    \mathcal{X}[\Omega] &= \frac{\Omega^2}{\sqrt{(\gamma_{\text{m}}^2 + \Omega^2)^2 + 2 \omega_{\text{m}}^2 (\gamma_{\text{m}}^2 - \Omega^2) + \omega^4}} \\
    &\approx \frac{\Omega^2}{(\Omega - \omega_{\text{m}})^2 + \gamma_{\text{m}}^2};
\end{split}
\end{equation}
and $2 \Xi$ is the additional phase accumulation relative to a free mass:
\begin{equation}
    \Xi[\Omega] = \frac{1}{2} \arctan \left( \frac{2 \kappa \Omega}{\gamma_{\text{m}}^2 + \omega_{\text{m}}^2 - \Omega^2} \right).
\end{equation}
We note that for a high quality factor mechanical oscillator far above mechanical resonance, where $\gamma_{\text{m}} \ll \omega_{\text{m}}$ and $\Omega \gg \omega_{\text{m}}$, we have $\mathcal{X}[\Omega] = 1$ and $\Xi[\Omega] = 0$, so \cref{eq:gwAntennaOutputQuads} reduces to the usual set of equations for an antenna in the free-mass regime.

We see that the gravitational wave signal is only coupled to the detector's output phase quadrature, $\hat{\alpha}_2^{\text{out}}$, and that this output field has five contributions. The first, proportional to $\hat{\alpha}_1^{\text{in}}[\Omega]$, is the radiation pressure noise of the laser light on the detector's mirrors, the second, proportional to $\hat{\alpha}_2^{\text{in}}[\Omega]$ is shot noise, and the third, proportional to $\hat{h}$ contains the gravitational wave signal and any associated fluctuations in this field. The final two, proportional to $\hat{\tilde{q}}_\text{in}$ and $\hat{\tilde{p}}_\text{in}$, come from thermal fluctuations in the test mass suspensions.

\subsection{Homodyne readout}

If a homodyne detector is used to measure the antenna's output phase quadrature, the detector's mean output is given by
\begin{equation}
    \langle\hat{\alpha}_2^\text{out}[\Omega]\rangle = e^{i \beta[\Omega]} \sqrt{2 \mathcal{K}[\Omega]} \frac{\langle\hat{h}[\Omega] \rangle}{h_\text{SQL}[\Omega]},
\end{equation}
and assuming uncorrelated noise modes, its output spectra are given by
\begin{subequations}
\begin{align}\label{eq:antennaOutputSpectra}
    \bar{S}_{\alpha \alpha}^{1,\text{out}}[\Omega] &= \bar{S}_{\alpha \alpha}^{1,\text{in}}[\Omega] \\
    \bar{S}_{\alpha \alpha}^{2,\text{out}}[\Omega] &= (\mathcal{K}[\Omega] \mathcal{X}[\Omega])^2 \bar{S}_{\alpha\alpha}^{1,\text{in}}[\Omega] + \bar{S}_{\alpha \alpha}^{2,\text{in}}[\Omega] \\
    & \qquad + \frac{2 \left|\mathcal{K}[\Omega]\right|}{h_\text{SQL}[\Omega]^2} \bar{S}_{h h}[\Omega] \\
    & \qquad + \mathcal{Y}[\Omega] \left(\frac{\gamma_{\text{m}}^2 + \Omega^2}{\omega_{\text{m}}} \bar{S}_{qq}^\text{in} + \omega_{\text{m}} \bar{S}_{pp}^\text{in} \right),
\end{align}
\end{subequations}
where we have defined
\begin{equation}
\begin{split}
    &\mathcal{Y}[\Omega] :=\\
    & \quad \frac{2 \hbar \kappa \gamma_{\text{m}} \bar{\alpha}^2 \omega_0^2}{m L^2 (\kappa^2 + \Omega^2)\left( (\gamma_{\text{m}}^2 + \Omega^2)^2 + 2 \omega_{\text{m}}^2(\gamma_{\text{m}}^2-\Omega^2) + \omega_{\text{m}}^4 \right)}.
\end{split}
\end{equation}
%
The first and second noise terms in the second line are the usual radiation pressure and shot noise terms. The third term is from the gravitational wave strain transduced to the detector's output phase quadrature. The final term is from dissipation in the test mass suspension.

\subsection{Photon counting detection}

Instead of using a homodyne detector to measure one output quadrature of a gravitational wave antenna, we could instead use a photon-number-resolving detector to measure the outgoing photon number operator~\cite{Blow90,Danilishin2012}
\begin{equation}
\begin{split}
    \hat{N}^\text{out}(t) &= \int\limits_{-\omega_0}^\infty \frac{\rmd\Omega}{2\pi} \int\limits_{-\omega_0}^\infty \frac{\rmd\Omega^\prime}{2\pi} \, \hat{e}^\text{out}[\Omega]^\dagger \, \hat{e}^\text{out}[\Omega^\prime] \, \rme^{\rmi(\Omega - \Omega^\prime)t} \\
    &\approx \int\limits_{-\infty}^\infty \frac{\rmd\Omega}{2\pi} \int\limits_{-\infty}^\infty \frac{\rmd\Omega^\prime}{2\pi} \, \hat{e}^\text{out}[\Omega]^\dagger \, \hat{e}^\text{out}[\Omega^\prime] \, \rme^{\rmi(\Omega - \Omega^\prime)t}.
\end{split}
\end{equation}
In the second line, we have used the fact that the optical frequency $\omega_0$ is very large to extend the lower bounds of the integrals to negative infinity.
It can be verified that $\left[\hat{N}^\text{out}(t), \hat{N}^\text{out}(t^\prime) \right] = 0$ so measurements of $\hat{N}^\text{out}$ can produce microscopically observable signals without introducing additional noise~\cite{Danilishin2012}. 

The output creation and annihilation operators $\hat{e}^\text{out}[\Omega]^\dagger$ and $\hat{e}^\text{out}[\Omega]$ are related to the output quadrature operators by
\begin{equation}
\begin{split}
    \hat{e}^\text{out}[\Omega] = \frac{\hat{\alpha}_1^\text{out}[\Omega] + i \hat{\alpha}_2^\text{out}[\Omega]}{\sqrt{2}} \\
    \hat{e}^\text{out}[\Omega]^\dagger = \frac{\hat{\alpha}_1^\text{out}[\Omega]^\dagger - i \hat{\alpha}_2^\text{out}[\Omega]^\dagger}{\sqrt{2}}.
\end{split}
\end{equation}
In the case of a resonant gravitational wave antenna, we have (using \cref{eq:antennaOutputSpectra})
\begin{equation}
\begin{split}
    N^\text{out}(t)&=\frac{1}{2}\int\limits_{-\infty}^\infty \frac{d\Omega}{2\pi} \left(\bar{S}_{\alpha\alpha}^{1,\text{out}}[\Omega] + \bar{S}_{\alpha\alpha}^{2,\text{out}}[\Omega] -1 \right) \\
    &= \frac{1}{2} \int\limits_{-\infty}^\infty \frac{d\Omega}{2\pi} \Bigg( \left(1 + (\mathcal{K}[\Omega] \mathcal{X}[\Omega])^2 \right) \bar{S}_{\alpha\alpha}^{1,\text{in}}[\Omega] \\
    & \qquad \qquad \qquad \quad + \bar{S}_{\alpha\alpha}^{2,\text{in}}[\Omega] - 1 \\
    &\qquad \qquad \qquad \quad + \mathcal{Y}[\Omega] \left(\frac{\gamma^2 + \Omega^2}{\omega_{\text{m}}} \bar{S}_{qq}^\text{in} + \omega_{\text{m}} \bar{S}_{pp}^\text{in} \right) \\
    & \qquad \qquad \qquad \quad + \frac{2 \left| \mathcal{K}[\Omega] \right|}{h_{\text{SQL}}[\Omega]^2} \bar{S}_{hh}[\Omega] \Bigg).
\end{split}
\end{equation}
All terms but the final one in this expression are present even in the absence of a gravitational wave. The final term comes from gravitons scattering photons to the detector's dark port.

We can define a notion of efficiency as the ratio of output photon click-rates $N^{\text{out}}$ to the 
rate $N^{\text{gw}}$ of gravitons emitted by the source (\cref{eq:Ngw}).
For a gravitational wave with carrier frequency $\Omega_0$, we find that the efficiency is
\begin{equation}\label{eq:etaLigo}
\begin{split}
    \eta_{\text{ifo}}[\Omega_0] &= \frac{N^{\text{out}}}{N^{\text{gw}}} \\
    &= \frac{\displaystyle\int\limits_{-\infty}^\infty \frac{\rmd{}\Omega}{2\pi} \frac{2 \left| \mathcal{K}[\Omega] \right|}{h_\text{SQL}[\Omega]^2} \bar{S}_{hh}[\Omega] }{\displaystyle\int\limits_{-\infty}^\infty \frac{\rmd{}\Omega}{2 \pi} \frac{R^2  c^3 \Omega_0}{ 5 \hbar G F} \bar{S}_{hh}[\Omega]} \\
    &\approx \frac{\displaystyle \frac{2 \left| \mathcal{K}[\Omega_0] \right|}{h_\text{SQL}[\Omega_0]^2} \int\limits_{-\infty}^\infty \frac{\rmd{}\Omega}{2\pi} \bar{S}_{hh}[\Omega] }{\displaystyle \frac{R^2  c^3 \Omega_0}{5 \hbar G F} \int\limits_{-\infty}^\infty \frac{\rmd{}\Omega}{2 \pi}  \bar{S}_{hh}[\Omega]} \\
    &= \frac{10 \hbar G F \left| \mathcal{K}[\Omega_0] \right|}{ c^3 R^2  \Omega_0 h_\text{SQL}[\Omega_0]^2},
\end{split}
\end{equation}
where we only consider output photons caused by the presence of a GW in calculating this ratio. More explicitly, the antenna's quantum efficiency, relative to the source, is given by
\begin{equation}
    \eta_{\text{ifo}}[\Omega_0] = \frac{5 \hbar G F \kappa \bar{\alpha}^2 \omega_0^2}{2 c^3 R^2 \Omega_0 (\kappa^2 + \Omega_0^2)}.
\end{equation}
Using $\bar{\alpha}^2 = (4L/c)P_{\text{cav}}/\hbar\omega_0$ for a circulating cavity power $P_{\text{cav}}$ (see \cref{fn:normalization}), we find for GW signals inside the cavity bandwidth ($\Omega_0 \ll \kappa$) that
\begin{widetext}
\begin{equation}\label{etaIfo}
    \eta_{\text{ifo}} \approx 7\times 10^{-74} \left(\frac{P_{\text{cav}}}{\text{1.0 MW}} \right) \left(\frac{\omega_0/2\pi}{\text{282 THz}} \right) \left(\frac{L}{\text{4 km}} \right) \left(\frac{\text{100 Mpc}}{R}\right)^2 \left(\frac{\text{60 Hz}}{\Omega_0/2\pi} \right) \left(\frac{\text{400 Hz}}{\kappa/2\pi} \right) \left(\frac{F}{1}\right).
\end{equation}
\end{widetext}
The fiducial values here are chosen to be similar to the designed parameters for 
Advanced LIGO~\cite{LIGOScientific:2014pky}.
Note again that this efficiency is the ratio of the graviton flux emitted by the source to the graviton flux absorbed by the detector.

\section{Continuous measurement of a bar's position}\label{app:continuousBarMeas}

Here we model the continuous measurement of a bar's position read out via a parametric transducer (as used for the Niobe detector \cite{blair94}), while the bar is simultaneously being driven by a gravitational wave. This mostly follows the analysis of continuous measurement of an oscillator's position presented in Ref.~\cite{ClerkRMP10}, in which the calculation is extended to include the interaction of the oscillator with a gravitational wave.

\subsection{Hamiltonian and equations of motion}
We model the bar as a mechanical oscillator of frequency $\omega_{\text{m}}$, mechanical decay rate $\gamma_{\text{m}}$, and position operator $\hat{x}$.
Although the parametric transducer for Niobe was operated in the microwave regime, we can model it similarly to an optical cavity with frequency $\omega_{\text{c}}$ and mode operator $\hat{d}$.
The total Hamiltonian for the system is
\begin{equation}
    \hat{H} = \hbar \omega_{\text{c}} \hat{d}^\dag \hat{d} + \hbar \omega_{\text{m}} \hat{b}^\dag \hat{b} +  \frac{\hbar \omega_{\text{c}} A_c\hat{x}}{x_{\mathrm{zpm}}}\hat{d}^\dag \hat{d} + \frac{ML}{\pi^2}\hat{\ddot{h}}\hat{x},
\end{equation}
where $\hat{x} =x_{\mathrm{zpm}}\left(\hat{b}+\hat{b}^{\dagger}\right)$, $x_{\mathrm{zpm}}=\sqrt{\hbar /\left(M \omega_m\right)}$ is zero point motion of the bar and $A_c$ is a coupling constant. The equations of motion for the transducer cavity modes and bar modes under this Hamiltonian are, respectively,
\begin{align}
    \dot{\hat{d}} &= -i \omega_{\mathrm{c}}\left(1 + \frac{A_c\hat{x}}{x_{\mathrm{zpm}}}\right) \hat{d} - \frac{\kappa}{2} \hat{a} - \sqrt{\kappa} \hat{d}_{\text{in}} \\
    \dot{\hat{b}} &= -i \omega_{\text{m}} \hat{b} - \frac{\gamma_{\text{m}}}{2} \hat{b} - \sqrt{\gamma_{\text{m}}} \hat{\rho} -i\frac{ML}{\pi^2\hbar}\,\ddot{\hat{h}}\,x_{\text{zpm}} -i\omega_c A_c \hat{d}^{\dagger} \hat{d},
\end{align}
where $\hat{h}$ is the operator for the gravitational field, and $\hat{\rho}$ is the mechanical vacuum noise satisfying $\left[\hat{\rho}(t), \hat{\rho}\left(t^{\prime}\right)\right]=\delta\left(t-t^{\prime}\right)$. If we now linearize the transducer cavity's creation and annihilation operators by the substitution $\hat{d} \rightarrow \bar{d} + \hat{d}$ for a coherent amplitude $\bar{d}$, and we neglect second-order terms, the equation of motion for the bar mode operator becomes
\begin{align}
    \dot{\hat{b}} &= -i \omega \hat{b} - \frac{\gamma_{\text{m}}}{2} \hat{b} - \sqrt{\gamma_{\text{m}}} \hat{\rho} -i\frac{ML}{\pi^2\hbar}\,\hat{\ddot{h}}\,x_{\text {zpm}} + g \left(\hat{d}^\dag - \hat{d} \right) \nonumber\\
   \label{mechoscillator}
    &= -i \omega \hat{b} - \frac{\gamma_{\text{m}}}{2} \hat{b} - \sqrt{\gamma_{\text{m}}} \hat{\rho} - i\frac{x_{\text{zpm}}}{\hbar}\left(\hat{F}_{\text{gw}} - \hat{F}\right).
\end{align}
Here, we have defined the following forces on the mechanical oscillator:
\begin{align}
    \hat{F}_{\text{gw}} &= \frac{ML}{\pi^2} \hat{\ddot{h}} \\
   \hat{F} &= - i \hbar g \left(\hat{d} -\hat{d}^\dag \right),
\end{align}
with the first being the force from the gravitational wave, and the second being the backaction force due to radiation pressure.
We now linearize the input modes via
\begin{equation}
    \hat{d}_{\text{in}}=\bar{d}_{\text{in}}+\hat{\xi}.
\end{equation}
Here, $\hat{\xi}$ are white noise vacuum fluctuations of the input field, where $[\hat{\xi}(t),\hat{\xi}(t') ] = \delta(t-t')$. Solving the equation of motion for the transducer cavity mode using a Fourier transform then leads to the following expressions of the cavity mode in terms of the position of the mirror and the photon shot noise (after simplifying for the $\Omega \ll \kappa$ limit):
\begin{equation}\label{doper}
    \hat{d}[\Omega] =\frac{2}{\kappa}\left(\sqrt{2}\,g\,\hat{z}[\Omega]-\sqrt{\kappa}\,\hat{\xi}[\Omega]\right),
\end{equation}
where $g \equiv -(2x_{\text{zpm}}/L_c)  \omega_{\mathrm{c}} \sqrt{\overline{\dot{N}}/\kappa}$, such that $\overline{\dot{N}}$ is the incoming photon flux to the photonic cavity, and $L_{\text{c}}$ is the length of the cavity. Therefore, the expression for the backaction force can be rewritten as:
\begin{equation}
    \hat{F}=-\frac{2i \hbar g}{ \sqrt{\kappa}}\left(\hat{\xi}^{\dagger} - \hat{\xi}\right).
\end{equation}
Now, we can use the input--output relation:
\begin{equation}
\hat{d}_{\text {out }}=\hat{d}_{\text {in }}+\sqrt{\kappa} \hat{d},
\end{equation}
which will enable us to write the homodyne quadrature variable
\begin{equation}
    \hat{I}=\hat{d}_{\text{out}} + \hat{d}_{\text{out}}^{\dagger}
\end{equation}
in terms of the position of the mechanical oscillator (which depends on the GW driving force)\,---\,since in \cref{doper}, the dependence of the cavity annihilation operator (which is included in the the quadrature output), depends on the position operator of the mechanical mirror. Therefore, in order to solve for $\hat{I}$, all that remains to be done is to solve for $\hat{z} = \hat{x}/x_{\text{zpm}}$. It is a straightforward exercise to solve \cref{mechoscillator} through a Fourier transform:
\begin{multline}
    \hat{z}[\Omega]  =  -\sqrt{\gamma_{\text{m}}}\left(\chi_{\mathrm{M}}[\Omega-\omega_{\text{m}}] \hat{\rho}[\Omega]+\chi_{\mathrm{M}}[\Omega+\omega_{\text{m}}] \hat{\rho}^{\dagger}[\Omega]\right) \\
    - \frac{i}{\hbar} x_{\mathrm{zpm}}\left(\chi_{\mathrm{M}}[\Omega-\omega_{\text{m}}]-\chi_{\mathrm{M}}[\Omega+\omega_{\text{m}}]\right) \left(\hat{F}[\Omega] + \hat{F}_{\text{gw}}[\Omega]\right),
    \label{eq:zbar}
\end{multline}
where \(\chi_{\text{M}}[\Omega - \omega_{\text{m}}] = 1/(-i(\Omega - \omega_{\text{m}}) + \gamma_{\text{m}}/2)\) is the mechanical susceptibility.
In \cref{eq:zbar}, the terms in $\rho$ and $\rho^\dagger$ are the mechanical noise of the oscillator, the term in $\hat{F}$ is the radiation pressure back-action, and the term in $\hat{F}_{\text{gw}}$ the response of the bar to the driving gravitational wave. 

\subsection{Homodyne detection}

Now, we can use this in the following expression for the output homodyne signal (which follows from the input--output relations):
\begin{equation}
\hat{I}=-\left(\hat{\xi}+\hat{\xi}^{\dagger}\right) + \frac{4g}{ \sqrt{\kappa}}\hat{z}.
\end{equation}
We can now proceed to compute the expectation value of the output homodyne signal for particular initial states of the GW (such as a coherent state of small amplitude). Most importantly, we see that the expected value of the output homodyne signal is directly proportional to the amplitude of the gravitational wave:
\begin{equation}
 \left\langle\hat{I}[\Omega]\right\rangle=-i \frac{4 g}{\sqrt{\kappa}} \sqrt{\frac{M}{\hbar \omega_{\text{m}}}} \frac{2 L \Omega_0^2}{\gamma_{\text{m}} \pi^2}\left\langle\hat{h}[\Omega]\right\rangle + \text{noise terms},
\end{equation}
Since the GW will be in a coherent state, the output will be proportional to its coherent state amplitude, regardless of how small that amplitude is.

\subsection{Photon counting detection}

Instead of using a homodyne detector to measure one output quadrature, we could instead use a photon-number-resolving detector to measure the outgoing photon number operator~\cite{Blow90,Danilishin2012}. Again, extending the integration bounds to negative infinity for large photonic frequencies,
\begin{equation}\label{nouthere}
    \hat{N}^\text{out}(t)=\int\limits_{-\infty}^\infty \frac{\rmd{}\Omega}{2\pi} \int\limits_{-\infty}^\infty \frac{\rmd{}\Omega^\prime}{2\pi} \, \hat{d}^\text{out}[\Omega]^\dagger \, \hat{d}^\text{out}[\Omega^\prime] \, \rme^{\rmi(\Omega - \Omega^\prime)t}.
\end{equation}
It can again be verified that $\left[\hat{N}^\text{out}(t), \hat{N}^\text{out}(t^\prime) \right] = 0$ so measurements of $\hat{N}^\text{out}$ can produce microscopically observable signals without introducing additional noise \cite{Danilishin2012}. Expanding \cref{nouthere}, we obtain 
\begin{equation}\label{nouthere2}
  \langle  \hat{N}^\text{out}(t)\rangle=\int\limits_{-\infty}^\infty \frac{\rmd{}\Omega}{2\pi}  \left(\bar{S}_{\xi^\dag \xi}[\Omega] + \frac{4g^2}{\kappa}\,\Bar{S}_{zz}[\Omega] \right).
\end{equation}
The first term in the integrand is the measurement imprecision, whereas the symmetrized double-sided cross-correlation spectrum (assuming uncorrelated noise modes between the three independent sub-systems corresponding to the gravitational radiation, electromagnetic cavity and phononic mode) is
\begin{widetext}
\begin{equation}\label{zz1}
\begin{split}
    \Bar{S}_{z z}[\Omega] =&  \gamma_{\text{m}}\left(\bigl|\chi[\Omega-\omega_{\text{m}}]\bigr|^2+ \chi[\Omega+\omega_{\text{m}}]\bigr|^2\right)\bar{S}_{\rho \rho}[\Omega] + (\gamma_{\text{m}}\bar{S}_{\rho \rho^\dag}[\Omega]\chi\left[\Omega-\omega_{\text{m}}\right]^\dag\chi\left[\Omega+\omega_{\text{m}}\right] + \mathrm{h.c.} ) \\
    &+\frac{x_{\mathrm{zpm}}^2}{\hbar^2}\bigl|\chi_{\mathrm{M}}[\Omega-\omega_{\text{m}}] -\chi_{\mathrm{M}}[\Omega+\omega_{\text{m}}]\bigr|^2 \left(\Bar{S}_{F F}[\Omega] + \frac{M^2L^2\Omega_0^4}{\pi^4}\Bar{S}_{h h}[\Omega] \right).
\end{split}
\end{equation}
If the mechanical mode is in a thermal state, we can take $\bar{S}_{\rho \rho}[\Omega] = \frac{1}{2} + \bar{n}$ and $\bar{S}_{\rho \rho^\dag}[\Omega] = 0$, yielding 
\begin{equation}\label{zz2}
\begin{split}
    \Bar{S}_{z z}[\Omega] =&  \gamma_{\text{m}}\left(\bigl|\chi[\Omega-\omega_{\text{m}}]\bigr|^2+ \chi[\Omega+\omega_{\text{m}}]\bigr|^2\right)\left( \frac{1}{2} + \bar{n} \right) \\
    & +\frac{x_{\mathrm{zpm}}^2}{\hbar^2}\bigl|\chi_{\mathrm{M}}[\Omega-\omega_{\text{m}}] - \chi_{\mathrm{M}}[\Omega+\omega_{\text{m}}]\bigr|^2 \left(\Bar{S}_{F F}[\Omega] + \frac{M^2L^2\Omega_0^4}{\pi^4}\Bar{S}_{h h}[\Omega] \right).
\end{split}
\end{equation}
\end{widetext}
The last term in \cref{zz2} demonstrates a response proportional to the square of the amplitude of the gravitational radiation. Therefore, even if there is a linear interaction between an optical cavity and the bar, a graviton-counter can be constructed by measuring the photon-flux of the photons that leak out of the cavity. The first term in \cref{zz2} arises due to the mechanical vacuum noise, and the second term due to the back-action of the photonic cavity on the mechanical oscillator.  

Therefore, detecting the photons that leak out of the cavity causes the bar to function as a square-law detector. In this case, if the gravitational wave is a coherent state of amplitude $\bar{a} \ll 1$, the output will be suppressed by $\mathcal{O}(\bar{a}^2)$, this contrasts the case of the homodyne measurement of the photons that leak out of the cavity, in which the signal is only suppressed by $\mathcal{O}(\bar{a})$.

We now compute the efficiency of the bar to emitted gravitons.
If we are in the regime for which the GW signal is well within the detection band, then $|\omega_{\text{m}} - \Omega| \ll \gamma_{\text{m}}$, and we have that the GW-driven bar motion, relative to the zero-point motion $x_{\text{zpm}}$, is $\bar{S}_{zz}^{\text{(GW)}}[\Omega] \approx (2x_{\text{zpm}}/\hbar \gamma_{\text{m}})^2 (M L \Omega_0^2 / \pi^2)^2 \bar{S}_{hh}[\Omega]$, so the number of GW-generated photons from the transducer cavity is
\begin{equation}
\begin{split}
  \langle N^{\text{out}} \rangle 
  &= \frac{4 g^2}{\kappa} \int\limits_{-\infty}^{\infty} \frac{\rmd\Omega}{2\pi} \bar{S}_{zz}^{\text{(GW)}}[\Omega] \\
  &= \frac{16 g^2}{\kappa} \frac{M L^2 \Omega_0^4}{\pi^4 \hbar \omega_{\text{m}} \gamma_{\text{m}}^2} \int\limits_{-\infty}^{\infty} \frac{\rmd\Omega}{2\pi} \bar{S}_{hh}[\Omega], \nonumber
\end{split}
\end{equation}
where we have taken the integration interval to infinity under the assumption the integrand only has support within the bandwidth of the bar, and we have used $x_{\text{zpm}} = \sqrt{\hbar/M\omega_{\text{m}}}$.
On the other hand, the number of incident gravitons is given by \cref{eq:Ngw}, and the ratio of the two gives
\begin{widetext}
\begin{align}\label{etaBar}
  \eta_{\text{bar}} &= \frac{96 g^2}{\kappa} \frac{M L^2 \Omega_0^3 G}{\pi^3 c^3 \omega_{\text{m}} \gamma_{\text{m}}^2 R^2} \\
    &\approx 1.2\times 10^{-61} \left(\frac{M}{\text{1000 kg}}\right) \left(\frac{\Omega_0/2\pi}{\text{1000 Hz}}\right)^2\left(\frac{1\times 10^{-5} \text{ Hz}}{\gamma_{\text{m}}/2\pi}\right)^2 \nonumber \left(\frac{L}{\text{3 m}}\right)^2\left(\frac{\text{100 Mpc}}{R}\right)^2\left(\frac{ g/2\pi}{\text{1000 Hz}}\right)^2\left(\frac{\text{100 Hz}}{\kappa/2\pi}\right)
\end{align}
\end{widetext}
where we have used bar detector parameters for Niobe~\cite{blair94}, and optomechanical readout coupling strengths and cavity decay rates used in its transducers~\cite{Cuthbertsonblair96}, we further assume a bar length $L = \text{3 m}$; we have also taken $\Omega_0 \approx \omega_{\text{m}}$. 
Just as for the interferometric case [\cref{etaIfo}], this efficiency is relative to the source; i.e., the ratio of the graviton flux emitted by the source to the graviton flux absorbed by the detector.

\section{Wait-time distribution}\label{app:waitTime}

While the spectral densities computed so far provide the ensemble average response of the detector, we now compute the waiting time for a detector click due to the absorption gravitons. This is necessary to confirm that for small coherent state amplitudes such that $\bar{a} \ll 1$, that the detector does not click to leading order in $\bar{a}$, further justifying that if the incident field can not be interpreted as consisting of a single graviton (by the absence of gravitons in the field, we interpret this as the probability of occupation of any Fock state higher than $\ket{n = 0}$ being negligible). 

In order to extend the computation of the waiting time distribution to the case of quantum fields, we will adapt the results derived in the electromagnetic case in Ref.~\cite{Carm93}, where the wait-time distribution was shown to be:
\begin{equation}
    w(\tau \mid t) \equiv \wp_2(t, t+\tau) / w_1(t),
\end{equation}
where
\begin{equation}
    \wp_2(t, t+\tau ) = \langle: \hat{I}(t+ \tau)e^{-\hat{\Omega}(t+\tau,t)} \hat{I}(t) :\rangle,
\end{equation}
\begin{equation}
    w_1(t) = \varepsilon \langle \hat{E}^{(-)}(t)\hat{E}^{(+)}(t) \rangle,
\end{equation}
and where $\hat{E}^{(-)}(t)$ and $\hat{E}^{(+)}(t)$ are the positive and negative frequency components, respectively:
\begin{subequations}
\begin{align}
    \hat{E}^{+}(t) &= i \int\limits_{-\infty}^{+\infty}  \rmd{}\Omega\left(\frac{\hbar \Omega}{4 \pi \varepsilon_0 c A}\right)^{1 / 2} \hat{a}[\Omega] \exp \left[-i \Omega t\right] \\
    \hat{E}^{-}(t) &= i \int\limits_{-\infty}^{+\infty}  \rmd{}\Omega\left(\frac{\hbar \Omega}{4 \pi \varepsilon_0 c A}\right)^{1 / 2} \hat{a}[\Omega]^\dagger \exp \left[+i \Omega t\right],
\end{align}
\end{subequations}
with absorption photon flux operator 
\begin{equation}
    \hat{I}(t)=\varepsilon \hat{E}^{(-)}(t) \hat{E}^{(+)}(t),
\end{equation}
and integrated flux operator as
\begin{equation}
\hat{\Omega}=\varepsilon \int\limits_t^{t+T} d t^{\prime} \hat{E}^{(-)}\left(t^{\prime}\right) \hat{E}^{(+)}\left(t^{\prime}\right).
\end{equation}
In this derivation for quantum fields, $\varepsilon$ is modified to be
\begin{equation}
    \varepsilon = \eta A \frac{2\varepsilon_0 c}{\hbar \Omega_0},
\end{equation}
with $\varepsilon_0 c$ being a constant that is added to ensure that $\hat{I}(t)$ has units of photon number flux. 

In order to construct such a wait-time distribution for the quantized gravitational field, we denote the following positive and negative frequency components of the quantized gravitational field:
\begin{subequations}
\begin{align}
    \hat{h}^+(t) &= \int\limits_{-\infty}^{\infty} \frac{\mathrm{d} \Omega}{2 \pi} \sqrt{\frac{8 \pi \hbar G}{c^3 \Omega_0 A}}a[\Omega] \mathrm{e}^{-\mathrm{i} (\Omega + \Omega_0)t}   \\
    \hat{h}^-(t) &= \int\limits_{-\infty}^{\infty} \frac{\mathrm{d} \Omega}{2 \pi} \sqrt{\frac{8 \pi \hbar G}{c^3 \Omega_0 A}} a[\Omega]^\dag \mathrm{e}^{\mathrm{i} (\Omega+ \Omega_0) t}.
\end{align}
\end{subequations}
This allows us to define the absorption-graviton-flux operator as
\begin{equation}
    \hat{I}(t) = \varepsilon \int\limits_{-\infty}^{\infty} \mathrm{d} \Omega \int\limits_{-\infty}^{\infty} \mathrm{d} \Omega^{\prime} \left( \frac{8 \hbar \pi G}{ c^3 |\Omega_0|  A} \right)\hat{a}[\Omega]^{\dagger} \hat{a}\left[\Omega^{\prime}\right] \mathrm{e}^{\mathrm{i}\left(\Omega-\Omega^{\prime}\right) t},
\end{equation}
where $\varepsilon = \frac{\eta A}{\hbar \Omega}\left(\frac{c^3\Omega^2}{32 \pi G}\right)$. Henceforth, we assume a constant coherent state amplitude $\bar{a}[\Omega]$ in the frequency domain, with source bandwidth $\Delta \Omega$, which gives the coherent state at the emitter in the time domain to be $\bar{a} \approx \Delta \Omega \, \bar{a}[\Omega]$; here we have also made the simplifying assumption that \((\Omega - \Omega')t \ll 1\). Therefore, we compute the expected number of gravitons absorbed to be
\begin{equation}
    \left\langle \hat{I}(t) \right\rangle = \varepsilon |\bar{a}|^2 \left( \frac{16\pi\hbar G}{ A c^3|\Omega_0|} \right) = \frac{\eta }{2 } |\bar{a}|^2.
\end{equation}
Again under the assumption that $(\Omega - \Omega^{\prime})t \ll 1$, we can simplify the integrated flux operator as
$\hat{\Omega} = \tau \hat{I}(t)$,
which enables a straightforward computation of the normal-ordered product. This leads directly to the wait-time distribution:
\begin{equation}\label{wtau}
w(\tau \mid t) = \frac{\eta}{4}|\bar{a}|^2\exp\left(-\tau \frac{\eta}{4}|\bar{a}|^2\right).
\end{equation}
Here, we see explicitly that as the coherent state amplitude $|\bar{a}|$ of the incoming radiation becomes much smaller than unity, the wait-time distribution vanishes to leading order in $|\bar{a}|^2$.

The efficiency that appears here is defined relative to the incident GW flux, i.e. it is the fraction 
of \emph{incident} GW flux that is converted to a signal at the output of the detector. This is 
different from the efficiencies relative to the source in \cref{etaIfo,etaBar}.
For a typical astrophysical source, the emitted flux is so large that the fraction of it 
incident on the detector is appreciable. To wit, assuming a coherent state emission, 
the coherent state amplitude at the detector can be estimated by equating the classical power flux with the energy flux of a single graviton. This gives: 
\begin{equation}
    |\bar{a}|^2 = \frac{\Omega_0 c^3\pi A }{32 G\hbar }h_0^2.
\end{equation}
Rewriting \cref{eq:etaLigo,etaBar} in terms of the cross-sectional area, we find
\begin{align}
    \eta_\text{ifo}[\Omega_0] &= \frac{8 \pi \hbar G \bar{\alpha}^2 \omega_0^2}{A c^3 \kappa \Omega_0} \\
    \eta_\text{bar}[\Omega_0] &= \frac{256 g^2 M L^2 \Omega_0^3 G}{3 \pi^2 c^3 \kappa \omega_\text{m} \gamma_\text{m}^2 A}
\end{align}
such that 
\begin{align}
    \eta_\text{ifo}[\Omega_0] |\bar{a}|^2 &=  \frac{\pi^2 \bar{\alpha}^2 \omega_0^2 h_0^2}{4 \kappa} \\
    \eta_\text{bar}[\Omega_0] |\bar{a}|^2 &= \frac{8 g^2 M L^2 \Omega_0^4 h_0^2}{3 \pi \hbar \kappa \omega_\text{m} \gamma_\text{m}^2}.
\end{align}
Note that the dependence of the quantization area in the efficiency is exactly canceled by the area factor in the flux,
so that their product $\eta |\bar{a}|^2$, and therefore the wait-time distribution, is independent of the 
details of quantization (as it must be). 

With the parameters used in \cref{etaIfo,etaBar} and a strain amplitude of $h_0 \sim 10^{-22}$, we obtain a graviton flux rate of $\eta_\text{ifo} |\bar{a}|^2 \sim 1\times 10^6\text{ s}^{-1}$ and $\eta_\text{bar} |\bar{a}|^2 \sim 1\times 10^{16}\text{ s}^{-1}$. 
This estimate for the bar is very optimistic and crude: we have assumed that the GW is monochromatic to within the $1 \times 10^{-5}\text{ Hz}$ mechanical linewidth of the bar's fundamental elastic mode; a more precise estimate in this case requires consideration of the actual GW template, and how that matches with the bar's impulse response.
On the other hand, the estimate for the case of an interferometric detector is realistic, since they are broadband.

\bibliography{main.bib}

\end{document}